\documentclass[12pt, draftclsnofoot, onecolumn]{IEEEtran}
\usepackage{pifont}
\usepackage{}
\usepackage[noadjust]{cite}
\usepackage{graphicx}
\usepackage{float}
\usepackage{amsmath}
\usepackage{enumerate}
\usepackage{amssymb}
\usepackage{fixltx2e}
\usepackage{color}
\usepackage{algorithm}
\usepackage{algorithmic}
\usepackage{amsmath}
\usepackage{multirow}
\usepackage{ragged2e}
\usepackage{booktabs}
\usepackage{subfigure}

\begin{document}

\pagenumbering{arabic}

\title{ Detection Interval for Diffusion Molecular Communication: How Long is Enough?}

\author{Xuan Chen, Miaowen Wen,~\emph{Senior Member,~IEEE,} Fei Ji,~\emph{Member,~IEEE,} \\
Yu Huang, Yuankun Tang,~Andrew W. Eckford,~\emph{Senior Member,~IEEE} \\

\thanks{X. Chen is with the School of Electronics and Information Engineering,
South China University of Technology, Guangzhou 510641, China, and also with the Department of Electrical Engineering and Computer Science, York University, Toronto M3J 1P3, Canada (Email:eechenxuan@mail.scut.edu.cn).

    F. Ji, M. Wen, and Y. Tang are with the School of Electronic and Information Engineering, South China University of Technology, Guangzhou 510640, China (Email: \{eefeiji, eemwwen\}@scut.edu.cn, eeyktang@mail.scut.edu.cn).

    Y. Huang is with the School of Electronics and Communication Engineering, Guangzhou University, Guangzhou 510006, China
    (Email: yuhuang@gzhu.edu.cn).

    A. W. Eckford is with the Department of Electrical Engineering and Computer Science, York University, Toronto M3J 1P3, Canada (Email: aeckford@yorku.ca).
}}

\maketitle
\vspace{-0.3cm}
\begin{abstract}

Molecular communication has a key role to play in future medical applications, including detecting, analyzing, and addressing infectious disease outbreaks. Overcoming inter-symbol interference (ISI) is one of the key challenges in the design of molecular communication systems. In this paper, we propose to optimize the detection interval to minimize the impact of ISI while ensuring the accurate detection of the transmitted information symbol, which is suitable for the absorbing and passive receivers. For tractability, based on the signal-to-interference difference (SID) and signal-to-interference-and-noise~amplitude~ratio (SINAR), we propose a modified-SINAR (mSINAR) to measure the bit error rate (BER) performance for the molecular communication system with a variable detection interval. Besides, we derive the optimal detection interval in closed form. Using simulation results, we show that the BER performance of our proposed mSINAR scheme is superior to the competing schemes, and achieves similar performance to optimal intervals found by exhaustive search.

% There is a recent surge of research interest in employing the internet of medical things (IoMT) to realize health monitoring remotely during the pandemic, in which a typical IoMT network is composed of sensors, local devices, and edge servers. An effective communication paradigm is essential for linking sensors all over the body and constructing an IoMT network. Among the existing communication methodologies, molecular communication (MC) is expected to achieve the above connection, especially for an IoMT built on nano bio-sensors. However, as a low-speed communication method based on diffusive property, overcoming inter-symbol interference (ISI) is one of the key challenges in the design of MC systems. In this paper, we propose to optimize the detection interval to minimize the impact of ISI while ensuring the acquisition of effective information. Our detection interval optimization applies to both the absorbing and passive receivers. For analysis, we propose a performance metric, namely modified signal-to-interference and noise amplitude ratio (mSINAR), to reduce the optimization complexity arising from the intractable bit error rate (BER). Based on this indicator, we derive an optimal detection interval in closed-form. Besides, BER simulation results validate the analysis and also show that the BER curve with the proposed mSINAR asymptotically approaches the best BER curve obtained from exhaustive search and significantly outperforms the existing schemes. Finally, simulation results also reveal that the optimal detection interval is convergent.

\end{abstract}


\date{\today}
\renewcommand{\baselinestretch}{1.2}
% \thispagestyle{empty} \maketitle \thispagestyle{empty}
%\newpage
\setcounter{page}{1}

\IEEEpeerreviewmaketitle
\newpage
%%%%%%%%%%%%%
\section{Introduction}

On a global scale, the COVID-19 pandemic is threatening the overall population health and economic well-being. Due to its rapid spread, how to offer fast disease detection along with “on-site” results is a major challenge to medical care. In this context, the Internet of Medical Things (IoMT), an extension and specialization of IoT in the medical industry, was proposed to realize remote health diagnosis and monitoring in the time of COVID-19 \cite{IoMT-1,IoMT-2,IoMT-3,IoMT-4}. In general, the IoMT network is composed of body sensors, local devices, and edge servers, in which the sensor is the basis of the entire network since it enables the collection of various health signals for the subsequent health assessment. Constrained by the volume and bio-compatibility of sensors, nano bio-sensors are considered a promising technology for the construction of IoMT. Besides, researchers also proposed using nano bio-sensors to facilitate various disease diagnoses and treatment, thus forming the Internet of Bio-Nano Things (IoBNT) \cite{IoNT-1,IoNT-2,IoNT-3}. At this point, we can regard the IoBNT as a special realization of IoMT in each individual.

To collect all monitored medical data from patients, the connection of sensors located in different tissues is a key issue to be solved for IoBNT or IoMT. In the literature, a series of communication technologies for biological nano-sensors, such as molecular communication (MC), electromagnetic communication, and acoustic communication, has been proposed. Unlike the conventional communication methods, MC is a bio-inspired communication paradigm, and its carrier of information is chemical signals. Accordingly, MC has the following properties: small size, energy efficiency, and excellent bio-compatibility \cite{MC}. Based on the above advantages, MC is believed to have the potential to connect larger groups of nano bio-sensors in IoBNT \cite{IoNT-1,IoNT-2,IoNT-3}. On the other hand, as an interdisciplinary communication mode, MC is capable of serving the biomedical field, in applications such as targeted drug delivery \cite{Drug-delivery-1}, \cite{Drug-delivery-2}, disease diagnosis \cite{Disease-diagnosis-1}, \cite{Disease-diagnosis-2}, and health monitoring \cite{Health-monotoring-1,Health-monotoring-2}. Moreover, during the outbreak of COVID-19 or other pandemics, MC can also be employed as an effective tool to model the spread of infections and diseases via aerosols \cite{COVID-1,COVID-2,COVID-3,COVID-4}. Against this background, MC can pave the way for the connection of sensors all over the body in IoMT.

Molecular communication via diffusion (MCvD) is a highly popular MC method that uses the free diffusion of carrier molecules to transfer information at the nanoscale. However, considering the diffusion channel, the inter-symbol interference (ISI) caused by the long delay of molecular diffusion is a great challenge for the MCvD system. The existence of ISI will adversely affect subsequent signal recovery and further lead to error propagation. At present, there are many proposed solutions to overcome ISI, mainly divided into the following three types: modulation-based \cite{modulation_based_1,modulation_based_2,modulation_based_3}; equalization-based \cite{equalization_based_1,equalization_based_2,equalization_based_3}; and channel-based, i.e., introducing an external factor, such as flow~\cite{Flow_ISI_mitigation_1}~or~enzyme~\cite{Enzyme_ISI_mitigation_1}, into the channel. Generally, these methods come at some increased cost, for example, more types of molecules/receptors \cite{modulation_based_1,modulation_based_3,equalization_based_3}; additional computational complexity \cite{equalization_based_1,equalization_based_2,equalization_based_3}; additional memory at nano-machines \cite{ISI_mitigation_memory}; and specific channel \cite{modulation_based_2,Flow_ISI_mitigation_1,Enzyme_ISI_mitigation_1}. Notably, the MCvD system requires simple implementation. However, the above prevailing ISI mitigation schemes inevitably increase the system complexity, though they can achieve better performance. Then a question naturally arises whether we can effectively mitigate the impact of ISI by simple yet effective techniques such as sampling.

% In this paragraph, we focus on detailing the existing ways to mitigate the ISI impact. Analyze the pros and cons, then introduce some ways from the detection interval to deal with the ISI.

In the literature, there are already some efforts toward eliminating the ISI on varying the detection interval. Particularly, these methods can be divided into three categories according to their characteristics: the shift-$\tau$ method; truncating the symbol duration in advance; and extracting a small portion of the symbol duration. First, the shift-$\tau$ method means that the receiver can shift its absorbing/counting interval backward by $\tau$ seconds so as to avoid the strong ISI region \cite{The_shift_method_1,The_shift_method_2,The_shift_method_3}. The optimal reception delay of the absorbing receiver was approximately derived in \cite{The_shift_method_1}, thus providing a reliable guide for receiver design. Second, truncating the symbol interval in advance means that the detection process will be terminated ahead of the end of a symbol period, as proposed in \cite{Truncation_1} for the MCvD system comprised of a transmitter, an absorbing receiver, and an interference source. The authors of \cite{Truncation_1} found a proper detection interval, which can effectively overcome the interference from unintended transmitters. Third, extracting a portion of the symbol duration means that the detection process will start late and end early. Compared with the previous schemes, this method considers the impact of ISI on the entire signal transmission process, rather than a specific time slot. Therefore, it can eliminate much of the interference caused by the ISI. In \cite{Extracting_1,Extracting_2}, via the simulation, Ntouni \emph{et al.} proved that this method can improve the system performance regardless of whether the flow is considered for the absorbing and passive receivers.

% However, this method only considers the influence of the front-end ISI while ignoring the rear-end interference. Meanwhile, it may require a higher design complexity, since the updated detection interval needs to span two different symbol durations.

% Nevertheless, the impact from ISI has not been analyzed when obtaining the detection interval, and this interval also cannot be solved directly by theoretical expressions.

% , yet the study was based on simulations without rigorous theoretical analysis.

Against the background, not only to overcome the incompleteness of the interference analysis in \cite{The_shift_method_1,The_shift_method_2,The_shift_method_3,Truncation_1} but also to provide rigorous theoretical analysis for how to achieve the optimal change of the detection interval, we propose to optimize the detection interval from the theoretical perspective in this paper. Different from \cite{Extracting_1,Extracting_2}, we will derive the analytic solution of the optimal detection interval, even when the number of transmitted molecules $\cal Q$ and the ISI length $L$ is not fixed. Specifically, the contributions of this paper are summarized as follows:
\begin{itemize}
    \item Given that the bit error rate (BER) expression in MCvD systems is generally intractable, we propose a performance metric, namely modified signal-to-interference~and~noise~amplitude~ratio (mSINAR), to formulate the objective functions and then quantify the impact of the detection interval on BER performance. Compared with the existing performance metrics investigated in \cite{Performance_metric}, mSIANR can be a good alternative to measure the BER performance for the MCvD system with a variable detection interval.
    \item Based on the proposed mSIANR, we derive a theoretical detection interval for the absorbing and passive receivers. Constrained by the high computational complexity arising from the summation of numerous ISI signals and noise, we split the objective function generated from mSINAR as two sub-problems to deduce the optimal detection interval for all considered $\cal Q$ and $L$. Besides, we provide valuable insight into the convergence of this interval with respect to $\cal Q$.
    \item We verify the accuracy of our analysis and compare the BER performance of the proposed mSINAR scheme with that of the existing schemes, such as the shift-$\tau$ method proposed in \cite{The_shift_method_1}, via computer simulations. Simulation results show that the BER curve with mSINAR significantly outperforms that of the existing schemes when the ISI length is greater than 1 and asymptotically approaches the best BER curve obtained from the exhaustive search.
\end{itemize}

The remainder of this paper is organized as follows. In Section~\ref{System Model and Problem Statement}, we review the fundamental of a typical MCvD system and introduce an optimization problem related to the detection interval. Section~\ref{Optimization Analysis} details the optimization analysis and derives the theoretical detection interval for the absorbing and passive receivers. The performance of the proposed mSINAR is evaluated in Section~\ref{Numerical results and analysis}, and finally, the conclusion is drawn in Section~\ref{Conclusion}.

\textit{Notation:} ${\cal B}\left( { \cdot \,, \cdot } \right)$, ${\cal N}\left( { \cdot \,, \cdot } \right)$, and ${\cal P}\left(  \cdot  \right)$ denote the Bernoulli distribution, Gaussian distribution, and Poisson distribution, respectively. $Q(\cdot)$ and ${\rm{erf}}\left(  \cdot  \right)$ are the Gaussian $Q$-function and the error function, respectively.

\section{System Model and Problem Statement} \label{System Model and Problem Statement}

\subsection{System Model}

\begin{figure}[t]
	\centering
	 \includegraphics[width=3.8in]{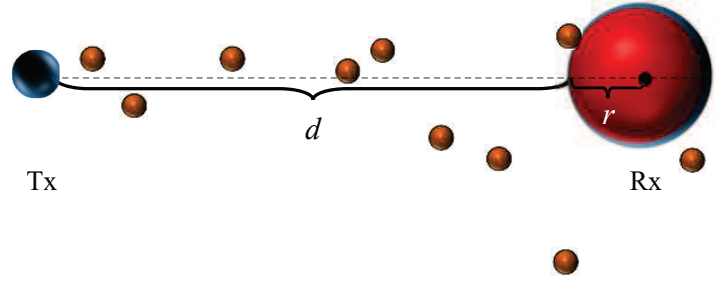}
	\caption{Schematic diagram of the MCvD system.}
	\label{figure-System-model}
\end{figure}

In this paper, we consider a typical MCvD system consisting of a point transmitter and a spherical receiver, which can be a fully absorbing or passive receiver. As can be seen from Fig.~\ref{figure-System-model}, the transmitter is placed in an unbounded 3-dimensional (3D) environment, in which the distance from the transmitter to the closest point of the receiver's surface is $d$ and the receiver's radius is $r$. For implementation, we assume that a transmitter can store (or generate) a certain number of information molecules with the diffusion coefficient $D$, and the corresponding receiver can recognize and distinguish these molecules. It is also assumed that the ON/OFF keying (OOK) modulation is applied to map the transmitted information to channel inputs. Specifically, the transmitter releases $\cal Q$ molecules to convey symbol ``1'', while releasing no molecules to convey symbol ``0''. Moreover, perfect time synchronization is assumed to ensure the detection interval scheduling. For clarity, in the following, we review the preliminary conceptual framework for MCvD systems, where the fully absorbing receiver and passive receiver are considered, respectively.

\subsubsection{Fundamentals of Absorbing Receiver}

Following Fick's law of diffusion, the probability of a molecule, released from a point transmitter at $t = 0$, reaching a spherical receiver at time $t$ is
\begin{align}\label{CIR_absorbing}
h(t)=\frac{r}{d+r} \frac{d}{\sqrt{4 \pi D t^{3}}} \exp \left(-\frac{d^{2}}{4 D t}\right).
\end{align}
Then we can express the expected fraction of molecules, absorbed by the receiver in $[t_1,t_2]$ with $t_1, t_2 \in [0, T_s]$ and $t_2 > t_1$, as
\begin{align}\label{CIR_absorbing_for_interval}
\hspace{-0.2cm}F_{\mathrm{ab}}\left(t_{1}, t_{2}\right)=\frac{r}{d+r}\left[\operatorname{erf}\left(\frac{d}{\sqrt{4 D t_{1}}}\right)-\operatorname{erf}\left(\frac{d}{\sqrt{4 D t_{2}}}\right)\right],
\end{align}
where $T_s$ denotes the symbol duration. Let us define $x_k$ and $Y_k$ as the $k$-th transmitted bit and the number of received molecules corresponding to the $k$-th transmission, respectively. Then we have
\begin{align}\label{Absorbing_molecules}
Y_k \sim \sum\limits_{i = 0}^L {\mathcal {B}\left( {{\cal Q}{x_{k - i}},{F_{ab}}\left( {{t_1} + iT_s,{t_2} + iT_s} \right)} \right)} ,
\end{align}
where $L$ is the ISI length. Following \cite{Truncation_1}, we use the Gaussian distribution rather than the Poisson distribution to approximate \eqref{Absorbing_molecules} as
\begin{align}\label{Approximated_Absorbing_molecules}
Y_k \sim \sum\limits_{i = 0}^L{\cal Q}{x_{k - i}}\mathcal{N}\left( {F_{ab}^i,F_{ab}^i\left( {1 - F_{ab}^i} \right)} \right),
\end{align}
where $F_{ab}^i={F_{ab}}\left( {{t_1} + iT_s,{t_2} + iT_s} \right)$. Moreover, we assume that the energy detection is used for the absorbing receiver.

From the above, we can write the average bit error probability as \cite{Extracting_1}
\begin{align}\label{P_e_AB_Gaussian}
{P_e} = \frac{1}{2} + \frac{1}{{{2^L}}}\sum\limits_{{\emph{\textbf{x}}_L} \in {{\cal M}^L}} {\left( {Q\left( {\frac{{\xi  - {\mu _0}\left( {{\emph{\textbf{x}}_L}} \right)}}{{{\sigma _0}\left( {{\emph{\textbf{x}}_L}} \right)}}} \right) - Q\left( {\frac{{\xi  - {\mu _1}\left( {{\emph{\textbf{x}}_L}} \right)}}{{{\sigma _1}\left( {{\emph{\textbf{x}}_L}} \right)}}} \right)} \right)} ,
% \hspace{-0.25cm}P_e \hspace{-0.08cm}=\hspace{-0.08cm} \frac{1}{2} \hspace{-0.05cm}+\hspace{-0.05cm} \frac{1}{{{2^L}}}{\sum _{{\emph{\textbf{x}}_L} \in {{\cal M}^L}}}\left(\hspace{-0.05cm} {Q\left( {\frac{{{{\xi}
% } - {\mu _0}}}{{{\sigma _0}}}} \right) \hspace{-0.05cm}-\hspace{-0.05cm} Q\left( {\frac{{{\xi} - {\mu _1}}}{{{\sigma _1}}}} \right)} \hspace{-0.05cm}\right),
\end{align}
where ${\xi}$ is the detection threshold; ${\mu_{x_k}}\left( {{\emph{\textbf{x}}_L}} \right)$ and $\sigma_{x_k}^2\left( {{\emph{\textbf{x}}_L}} \right)$ are the expectation and variance of $Y_k$ when ${x_k} \in \{0, 1\}$, expressed as
\begin{align}
{\mu _{{x_k}}}\left( {{\emph{\textbf{x}}_L}} \right) = \sum\limits_{i = 0}^L {{\cal Q}{x_{k - i}}F_{ab}^i},~\sigma _{{x_k}}^2\left( {{\emph{\textbf{x}}_L}} \right) = \sum\limits_{i = 0}^L {{\cal Q}{x_{k - i}}F_{ab}^i\left( {1 - F_{ab}^i} \right)},
\end{align}
and ${\emph{\textbf{x}}_L}$ is the ISI sequence with length $L$. For the $k$-th transmitted bit, ${\emph{\textbf{x}}_L}= \left\{ {{x_{k - L}}, \cdots ,{x_{k - 2}},{x_{k - 1}}} \right\}$ and ${x_{k-i}} \in {\cal M}$ with $i \in \left\{ {1, \cdots ,L} \right\}$ and ${\cal M} \in \{0, 1\}$. For clarity, in the sequel, we will ignore $\left( {{\emph{\textbf{x}}_L}} \right)$ for ${\mu _{{x_k}}}\left( {{\emph{\textbf{x}}_L}} \right)$ and $\sigma _{{x_k}}^2\left( {{\emph{\textbf{x}}_L}} \right)$. Besides, in \eqref{P_e_AB_Gaussian}, we also assume that the probabilities of transmitting bit ``0'' and bit ``1'' are ${1 \mathord{\left/{\vphantom {1 2}}\right.\kern-\nulldelimiterspace} 2}$.

\subsubsection{Fundamentals of Passive Receiver}

Similar to the absorbing receiver, we can define when $\frac{r}{r+d} <0.15$, the probability of observing a given molecule, emitted from the point transmitter at $t = 0$, inside $V$ at time $t$ as \cite{passive_receiver_probability}
\begin{align}\label{Passive_CIR}
p\left( t \right) = \frac{V}{{{{\left( {4\pi Dt} \right)}^{3/2}}}}\exp \left( { - \frac{{{{\left( {d + r} \right)}^2}}}{{4Dt}}} \right),
\end{align}
where $V$ is the volume of the passive receiver. First, we assume that $N$ samples are taken by the receiver at a symbol duration given by $f(n) \in \left[ {0,{T_s}} \right]$ where $n = 0,1, 2,\cdots,N$, and that they are equally summed up before the single threshold detection. The number of received molecules corresponding to the $n$-th sample taken within the $k$-th symbol duration can be expressed as
\begin{align}\label{Counting_molecules}
Y_{n,k} \sim {\cal P}\left( {{\Lambda _{n,k}} = \sum\limits_{i = 0}^L {{\cal Q}}{{x_{k - i}}p\left( {f\left( n \right) + i{T_s}} \right)} } \right),
\end{align}
where $x_k$ is the $k$-th transmitted bit and ${\Lambda _{n,k}}$ is defined as the average of the number of received molecules in this sample. Following most of the research works that have been conducted so far, we also sample the signals at equal intervals, i.e., $f(n) = nt_s$ and $t_s = T_s/N$ where $n = 0, 1, 2,\cdots,N$. Moreover, for clarity, we define $p_{n,i} = p\left( {f\left( n \right) + i{T_s}} \right)$. Considering the fact that if the Poisson parameter ${\Lambda _{n,k}}$ is sufficiently large, e.g., ${\Lambda _{n,k}} > 20$, we use the Gaussian distribution to approximate \eqref{Counting_molecules} as
\begin{align}\label{Approximated_Counting_molecules}
Y_{n,k} \sim {\cal N}\left( {{\Lambda _{n,k}},{\Lambda _{n,k}}} \right).
\end{align}
At this point, the average error bit probability for the passive receiver can be expressed as \eqref{P_e_AB_Gaussian}, where $\mu_{x_k}$ and $\sigma_{x_k}$ are replaced by $\sum\limits_{n = n_1}^{n_2}{{\mu _{x_k}}\left( n \right)}$ and ${\sqrt {\sum\limits_{n = n_1}^{n_2}{{\mu _{x_k}}\left( n \right)} } }$ for ${x_k}\in \left\{ {0,1} \right\}$; $n_1$ and $n_2$ denote the first sample and the last sample, respectively, with ${0 \le {n_1},{n_2} \le N}$. Further, ${\mu_{x_k}}\left( n \right)$ is the expectation of $Y_{n,k}$. Therefore, we have
\begin{align}
{\mu _{{x_k}}}\left( n \right) = \sum\limits_{i = 0}^L {{\cal Q}{x_{k - i}}p_{n,i}}.
\end{align}

\subsection{Problem Statement}

In most of the research works related to the MCvD system, it is generally assumed that the detection interval is equal to the transmission
symbol duration. However, some researchers have noted that the variable detection interval could have a great potential to improve the BER performance, especially when the ISI is relatively serious. Based on this fact, we propose to determine the optimal detection interval from the theoretical perspective, mathematically expressed as
\begin{align}\label{objective_AB}
{\left[ {{t_1},{t_2}} \right]^*}&=\hspace{-0.1cm} \mathop {\arg \min }\limits_{0 \le {t_1},{t_2} \le T_s} {P_e} \nonumber\\
&=\mathop {\arg \min }\limits_{0 \le {t_1},{t_2} \le T_s} \sum_{{\emph{\textbf{x}}_L} \in {{\cal M}^L}} \left[ {Q\left( {\frac{{\xi  - {\mu _0}}}{{{\sigma _0}}}} \right) - Q\left( {\frac{{\xi  - {\mu _1}}}{{{\sigma _1}}}} \right)} \right],
% &=\mathop {\arg \min }\limits_{0 \le {t_1},{t_2} \le T_s} \sum_{{\emph{\textbf{x}}_L} \in {{\cal M}^L}} {{\frac{{Q\left( {\frac{{{\xi} - {\mu _0}}}{{{\sigma _0}}}} \right) - Q\left( {\frac{{{\xi} \hspace{-0.03cm}-\hspace{-0.03cm} {\mu _1}}}{{{\sigma _1}}}} \right)}}{{{2^L}}}}}.
\end{align}
where the absorbing receiver is considered. The above function is also suitable for the passive receiver, in which ${\left[ {{t_1},{t_2}} \right]^*}$ is replaced by ${\left[ {{n_1},{n_2}} \right]^*}$ with $0 \le {n_1},{n_2} \le N$.
% In this subsection, in order to reduce the complexity of optimizing $[t_1, t_2]$, we first introduce the existing performance metrics as alternatives to calculate the optimal $[t_1, t_2]$. Further, based on the above metrics, we propose a revised performance indicator to symbolize the BER trend such that the best BER performance can be easily and precisely attained.
From \eqref{objective_AB}, it is difficult to directly obtain the optimal ${\left[ {{t_1},{t_2}} \right]}$ except with the aid of the exhaustive search. Thus, it is essential to determine how to obtain an exact or approximate ${\left[ {{t_1},{t_2}} \right]^*}$. In the literature, using the different performance metrics as alternatives of BER has been proposed in the MC field \cite{Performance_metric}. For clarity, in the following, we will take the absorbing receiver as an example to review these performance metrics:

1) Signal-to-interference ratio (SIR), which is the ratio of the amount of desired signal molecules to interference molecules:
\begin{align}\label{SIR_definition}
\textrm{{{SIR}}} = \frac{{{\cal Q}F_{ab}^0}}{{\sum\limits_{k = 1}^L {\cal Q} F_{ab}^k}}.
\end{align}

2) Signal-to-interference difference (SID), which is the difference between the number of desired signal molecules and interference molecules:
\begin{align}\label{SID_definition}
\textrm{{{SID}}} = {\cal Q}F_{ab}^0 - \sum\limits_{k = 1}^L {\cal Q} F_{ab}^k.
\end{align}

3) Signal-to-interference~and~noise~amplitude~ratio (SINAR), which means in addition to the interference sources, the noise source is also included in the calculation, having a form of
\begin{align}\label{SINAR}
\textrm{{SINAR}} = \frac{{{\cal Q}F_{ab}^0}}{{\sum\limits_{k = 1}^L {\cal Q} F_{ab}^k + \sum\limits_{k = 0}^L {\sqrt {{\cal Q}F_{ab}^k\left( {1 - F_{ab}^k} \right)} } }}.
\end{align}
Notice that the strongest ISI has been considered for all performance metrics, i.e., the ISI signals are assumed to be as ``$\underbrace {1 \cdots 1 \cdots 1}_L$''. Moreover, we can see that only SINAR involves the impact from the noise (i.e., the variation of signals). Therefore, SINAR is more realistic than the other metrics. Besides, SIR can be regarded as a limit of SINAR, expressed as follows
\begin{align}\label{SINAR-SIR}
 \mathop {\lim }\limits_{{\cal Q} \to \infty } {\mathop{\textrm{{SINAR}}}\nolimits}  &= \mathop {\lim }\limits_{{\cal Q} \to \infty } \frac{{F_{ab}^0}}{{\sum\limits_{k = 1}^L {F_{ab}^k}  + \sum\limits_{k = 0}^L {\sqrt {\frac{{F_{ab}^k\left( {1 - F_{ab}^k} \right)}}{\cal Q}} } }} \nonumber\\
  &= {\textrm{{SIR}}}.
\end{align}
Specially, in cases where the signal is not strong, but the unexpected ISI signal is very weak (i.e., nearly zero), SIR has extremely high values, which may cause an inaccurate analysis. As for SID, it is formulated by the difference between the numerator and denominator of SIR such that it can avoid the above extreme case. Therefore, in \cite{Performance_metric}, SID is considered to be a more reliable performance metric than SIR. Besides, according to the previous research, maximizing SINAR and SID can yield close-to-optimal results in terms of minimizing BER.

In view of the foregoing, it seems feasible to use SINAR and SID to calculate the approximate ${\left[ {{t_1},{t_2}} \right]^*}$. However, it is worth mentioning that this is the first time to use variable ${\left[ {{t_1},{t_2}} \right]}$ with ${t_2} - {t_1} \ne {T_s}$ in these indicators, so their applicability is still an open issue. First, we notice that in SINAR, when ${F_{ab}^k}$ with $k = 0,1, \cdots L$ varies with the change of the candidate ${\left[ {{t_1},{t_2}} \right]}$, the peak value of SINAR is always nearby $t_{\max}$ or saying that ${\left[ {{t_1},{t_2}} \right]^*}$ eventually converges near $t_{\max}$, as $\cal Q$ increases, where ${t_{\max }}$ represents the peak time for the molecule concentration when an impulse of molecules is emitted at $t = 0$, formulated as $t_{\max}=\frac{\left(d+r\right)^{2}}{6 D}$ for the absorbing receiver. This phenomenon can be explained as follows: since the influence of noise is inversely proportional to $\cal Q$, all interference (i.e., the denominator of SINAR) can be approximately regarded as linearly related to the desired signal, when $\cal Q$ is large enough; while in this case, the peak value of the ratio between the expected signals and the unexpected part should be near $t_{\max}$, due to the property of the diffusion channel. Yet, from the passive receiver, we know that the threshold detection with the multi-sampling is generally superior to that of the single sampling, which suggests that SINAR may not provide an accurate quantification for BER with a variable ${\left[ {{t_1},{t_2}} \right]}$. Moreover, as for SID, this metric assumes the influence of noise as 0, resulting in the deduced ${\left[ {{t_1},{t_2}} \right]^*}$ being independent of $\cal Q$. Clearly, this assumption is unreasonable, since only the noise with an infinite $\cal Q$ may go to zero.

Based on the above metrics, we propose a modified performance metric, namely modified-SINAR (mSINAR), which can be mathematically defined as
\begin{align}\label{mSINAR}
{\textrm{mSINAR}} = \frac{{\frac{1}{2}F_{ab}^0}}{{\sum\limits_{k = 1}^L {\frac{1}{2}} F_{ab}^k + \sum\limits_{k = 0}^L {\sqrt {\frac{{F_{ab}^k\left( {1 - F_{ab}^k} \right)}}{{2{\cal Q}}}} } }}.
\end{align}
First, we can see that mSINAR has a similar expression with SINAR except with the introduction of the transmission probability, where it is assumed that the probabilities of transmitting bit ``0'' and bit ``1'' are equivalent. Moreover, given the fact that ${\left[ {{t_1},{t_2}} \right]^*}$ obtained from SIANR is gradually closer to $t_{\max}$ with an increasing $\cal Q$, we need to define a valid range for mSINAR. It is clear from \eqref{mSINAR} that the impact from the noise will play a dominant role when $\cal Q$ is relatively small, suggesting that the initial value of mSINAR infinitely approaches 0 with an infinitesimal $\cal Q$; while as $\cal Q$ goes larger, the expected signal progressively shows its advantage in mSINAR, causing the increase of mSINAR until it reaches or exceeds 1. Therefore, we have $0 < {\textrm{mSINAR}}  \le {{1}}$. Furthermore, we regular the usage strategy of mSINAR for the derivation of ${\left[ {{t_1},{t_2}} \right]^*}$: when mSINAR is located in the valid range, we use mSINAR to derive the optimal detection interval; while as for ${\text{mSINAR}} > 1$, we continue to employ the ${\left[ {{t_1},{t_2}} \right]^*}$ obtained from this metric with $\textrm{mSINAR} = 1$. In the sequel, we will employ the proposed mSINAR to simplify the objective function in \eqref{objective_AB} and then optimize the detection interval.

Moreover, we also modify the SID to simplify the derivation of the optimal $ {\left[ {{t_1},{t_2}} \right]}$ in the next section. Specifically, we convert mSINAR to modified-SID (mSID), which is defined as the difference between the number of desired molecules and undesired molecules contributing by the ISI signal and noise, mathematically expressed as
\begin{align}\label{mSID_definition}
{\textrm{mSID}} = F_{ab}^0 - \sum\limits_{k = 1}^L {F_{ab}^k}  - \sqrt {\frac{2}{\cal Q}} \sum\limits_{k = 0}^L {\sqrt {F_{ab}^k\left( {1 - F_{ab}^k} \right)} } .
\end{align}
Compared with the original SID, mSID further includes the impact from the noise.

% , i.e., $0 < {\textrm{mSINAR}}  \le {{1}}$. It is clear from \eqref{mSINAR} that the impact from the counting noise will play a dominant role when $\cal Q$ is relatively small, suggesting that the initial value of mSINAR infinitely approaches 0 with an infinitesimal $\cal Q$; while as $\cal Q$ goes larger, the expected signal progressively shows its advantage in mSINAR, causing the increase of mSINAR until it reaches or exceeds~1. Per the aforementioned, we propose the using strategy of mSINAR for the derivation of ${\left[ {{t_1},{t_2}} \right]^*}$ as follows: when mSINAR is located in the valid range, we will use mSINAR to derive the optimal detection interval; while as for ${\text{mSINAR}} > 1$, we continue to employ the ${\left[ {{t_1},{t_2}} \right]^*}$ obtained from this metric with $\textrm{mSINAR} = 1$.

\section{Optimization Analysis} \label{Optimization Analysis}

In this section, we use the above mSINAR-based approximation methods to solve the optimal detection interval, where a fully absorbing receiver and a passive receiver are both considered.

\subsection{Detection Interval Optimization for Absorbing Receiver}

First, we discuss how to obtain ${\left[ {{t_1},{t_2}} \right]^*}$ for the absorbing receiver when mSINAR is used. Based on the usage strategy of mSINAR described previously, the objective function in \eqref{objective_AB} can be rewritten as
\begin{align}\label{mSINAR_objective}
 {\left[ {{t_1},{t_2}} \right]^*} &= \mathop {\arg \min }\limits_{0 \le {t_1},{t_2} \le {T_s}}{P_e} \nonumber\\
 &\approx \left\{ \begin{array}{l}
 \mathop {\arg \max }\limits_{0 \le {t_1},{t_2} \le {T_s}} \frac{{F_{ab}^0}}{{\sum\limits_{k = 1}^L {F_{ab}^k}  + \sum\limits_{k = 0}^L {\sqrt {\frac{{2F_{ab}^k\left( {1 - F_{ab}^k} \right)}}{{\cal Q}}} } }},~\textrm{if}~0 < {\cal Q} < \hat {\cal Q}  \\
 \mathop {\arg \max }\limits_{0 \le {t_1},{t_2} \le {T_s}} \frac{{F_{ab}^0}}{{\sum\limits_{k = 1}^L {F_{ab}^k}  + \sum\limits_{k = 0}^L {\sqrt {\frac{{2F_{ab}^k\left( {1 - F_{ab}^k} \right)}}{{\hat {\cal Q}}}} } }},~\textrm{if}~{\cal Q} \ge \hat {\cal Q}
 \end{array} \right. .
\end{align}
Here, $\hat {\cal Q} = \left\lceil {2{{\left( {\frac{{\sum\limits_{k = 0}^L {\sqrt {F_{ab}^k\left( {1 - F_{ab}^k} \right)} } }}{{F_{ab}^0 - \sum\limits_{k = 1}^L {F_{ab}^k} }}} \right)}^2}} \right\rceil$
% \begin{align}\label{special_Q}
% \hat {\cal Q} = \left\lceil {2{{\left( {\frac{{\sum\limits_{k = 0}^L {\sqrt {F_{ab}^k\left( {1 - F_{ab}^k} \right)} } }}{{F_{ab}^0 - \sum\limits_{k = 1}^L {F_{ab}^k} }}} \right)}^2}} \right\rceil.
% \end{align}
, calculated from \eqref{mSINAR} when assuming mSINAR = 1, where ${\left[ {{t_1},{t_2}} \right]^*}$ is also assumed to be used. According to the valid range of mSINAR, we split \eqref{mSINAR_objective} into the following two sub-problems to derive ${\left[ {{t_1},{t_2}} \right]^*}$, respectively.

\subsubsection{$0 < {\cal Q} < \hat {\cal Q}$} First, provided that $F_{ab}^k$ is a decreasing function with respect to $k$ and the value of $F_{ab}^k$ is relatively small, we approximate $\left( {1 - F_{ab}^k} \right)$ as 1. Thus, \eqref{mSINAR_objective} can be re-written as
\begin{align}\label{mSINAR_objective_first_subproblem}
{\left[ {{t_1},{t_2}} \right]^*} \approx \mathop {\arg \max }\limits_{0 \le {t_1},{t_2} \le {T_s}} \frac{{F_{ab}^0}}{{\sum\limits_{k = 1}^L {F_{ab}^k}  + \sqrt {\frac{2}{\cal Q}} \sum\limits_{k = 0}^L {\sqrt {F_{ab}^k} } }},~0 < {\cal Q} < \hat {\cal Q} .
\end{align}
Considering that \eqref{mSINAR_objective_first_subproblem} is related to $t_1^*$ and $t_2^*$, we can resort to the second partial derivative test to find a maximum value of mSINAR. Taking the first derivative of \eqref{mSINAR_objective_first_subproblem} with respect to $t_1$ and $t_2$, we can obtain
\begin{align}\label{mSINAR_derivative}
\frac{{\partial \textrm{mSINAR}}}{{\partial {t_i}}} =&{\left( { - 1} \right)^i}\left[ h\left( {{t_i}} \right){\left\{ {\sum\limits_{k = 1}^L {\int_{{t_1}}^{{t_2}} {h\left( {k{T_s} + t} \right)dt}  + } \sum\limits_{k = 0}^L {\sqrt {\frac{2}{\cal Q}} \sqrt {\int_{{t_1}}^{{t_2}} {h\left( {k{T_s} + t} \right)dt} } } } \right\}} \right.\nonumber\\
&\left.- \int_{{t_1}}^{{t_2}} {h\left( t \right)dt} {\left\{ {\sum\limits_{k = 1}^L {h\left( {k{T_s} + {t_i}} \right)}  + \sum\limits_{k = 0}^L {\frac{{h\left( {k{T_s} + {t_i}} \right)}}{{\sqrt {2{\cal Q}\int_{{t_1}}^{{t_2}} {h\left( {k{T_s} + t} \right)dt} } }}} } \right\}} \right] \nonumber\\
&\div {\left( {\sum\limits_{k = 1}^L {\int_{{t_1}}^{{t_2}} {h\left( {k{T_s} + t} \right)dt} }  + \sqrt {\frac{2}{\cal Q}} \sum\limits_{k = 0}^L {\sqrt {\int_{{t_1}}^{{t_2}} {h\left( {k{T_s} + t} \right)dt} } } } \right)^2},~i = 1,2.
\end{align}
From the above equation, we find that it is almost unavailable to solve the ${\left[ {{t_1},{t_2}} \right]^*}$ by making $\frac{{\partial \textrm{mSINAR}}}{{\partial {t_i}}} = 0$. Therefore, we decide to seek an alternative optimization objective function for \eqref{mSINAR_objective_first_subproblem}. First, it is worth noting that compared with $t_2^*$, $t_1^*$ is almost slightly affected by the noise, due to $t_1^* \in \left[ {0,{t_{\max }}} \right]$ and ${t_{\max }} \ll T_s$. Based on this assumption, we decompose the joint optimization problem of
${\left[ {{t_1},{t_2}} \right]^*}$ as two independent problems. For $t_1^*$, we resort to SID to solve it when assuming $t_2^* = T_s$, where the impact from the noise has been ignored. At this point, we have the following problem formulation
\begin{align}\label{SID_objective_t_1}
t_1^* &= \arg \min {P_e} \nonumber\\
&\approx \mathop {\arg \max }\limits_{0 \le {t_1} \le {t_{\max }}} \int_{{t_1}}^{T_s} {\left[ {h\left( t \right) - \sum\limits_{k = 1}^L {h\left( {k{T_s} + t} \right)} } \right]dt}.
\end{align}
As for $t_2^*$, we can re-write \eqref{mSINAR_objective_first_subproblem} as
\begin{align}\label{mSINAR_objective_t_2}
 t_2^* \approx \mathop {\arg \max }\limits_{{t_{\max }} < {t_2} \le {T_s}} \frac{{\int_{t_1^*}^{{t_2}} {h\left( t \right)dt} }}{{f\left( {{t_2}} \right)}},
\end{align}
where for clarity, let $f\left( {{t_2}} \right) = \sum\limits_{k = 1}^L {\int_{t_1^*}^{{t_2}} {h\left( {k{T_s} + {t}} \right)dt} }  + \sqrt {\frac{2}{\cal Q}} \sum\limits_{k = 0}^L {\sqrt {\int_{t_1^*}^{{t_2}} {h\left( {k{T_s} + {t}} \right)dt} } }$. However, after taking the first and second derivative of $\frac{{\int_{t_1^*}^{{t_2}} {h\left( t \right)dt} }}{{f\left( {{t_2}} \right)}}$, we find that it is highly involved to directly determine the sign of the second derivative of $\frac{{\int_{t_1^*}^{{t_2}} {h\left( t \right)dt} }}{{f\left( {{t_2}} \right)}}$, suggesting that we cannot resort to the second partial derivative test to find a $t_2^*$ for \eqref{mSINAR_objective_first_subproblem}.
Hence, we decide to obtain a close-to-optimal ${t_2}$ with a closed-form expression by using the similar method in \eqref{SID_objective_t_1}, i.e., assuming that there is no impact from the noise. Here, combining \eqref{SID_objective_t_1}, we can re-write \eqref{mSINAR_objective_first_subproblem} with $0 < {\cal Q} < \hat {\cal Q}$ as
\begin{align}\label{SID_objective}
{\left[ {{t_1},{t_2}} \right]^*} &= \arg \min {P_e} \nonumber\\
&\approx \mathop {\arg \max }\limits_{0 \le {t_1},{t_2} \le {T_s}} \int_{{t_1}}^{{t_2}} {\left[ {h\left( t \right) - \sum\limits_{k = 1}^L {h\left( {k{T_s} + t} \right)} } \right]dt} .
\end{align}
In \eqref{SID_objective}, we find that $h\left( t \right)$ is a function that rises sharply and falls slowly, while $ \sum\limits_{k = 1}^L {h\left( {kT_s+t} \right)}$ is a decreasing function. This means that \eqref{SID_objective} is equivalent to collecting all of intervals satisfying $ h{\left( t \right) - \sum\limits_{k = 1}^L {h\left( {k{T_s} + t} \right)}} \ge 0$, and we have
\begin{align}\label{find_interval_AB}
h{\left( t_i^* \right) - \sum\limits_{k = 1}^L {h\left( {k{T_s} + t_i^*} \right)} } = 0 , i=1,2.
\end{align}
However, note that \eqref{SID_objective} and \eqref{find_interval_AB} are valid when there is always a $\left[ {{t_1},{t_2}} \right]$ to satisfy
\begin{align}
\int_{{t_1}}^{{t_2}} {\left[ {h\left( t \right) - \sum\limits_{k = 1}^L {h\left( {k{T_s} + t} \right)} } \right]dt}  > 0, \nonumber
\end{align}
since when the above condition is false, the monotonicity of $h\left( t \right)$ and $ \sum\limits_{k = 1}^L {h\left( {kT_s+t} \right)}$ indicates that the solution of \eqref{SID_objective} is either $t_1^* = t_2^* = {t_{\max }}$ or no solution. Furthermore, we can see that it is still relatively tough to directly solve \eqref{find_interval_AB}, due to the summation caused by $L$ ISI signals. Therefore, according to different $L$, we decompose \eqref{find_interval_AB} as two cases, i.e., $L=1$ and $L>1$, to calculate the approximate $ {\left[ {{t_1},{t_2}} \right]^*}$, respectively. Besides, it is clear from \eqref{find_interval_AB} that ${\left[ {{t_1},{t_2}} \right]^*}$ solved by SID is independent of $\cal Q$, since the impact of noise has been neglected in SID. For clarity, the derived $ {\left[ {{t_1},{t_2}} \right]^*}$ for all possible $L$ is described below.

\textbf{\emph{Proposition 1:}} Assuming that the approximation for \eqref{find_interval_AB} with $L = 1$ is exact, the solution of \eqref{find_interval_AB} with $L = 1$ and $0 < {\cal Q} < \hat {\cal Q}$ can be solved as,
\begin{align}\label{final_t1_t2_solution_absorbing_SID_L_1}
{\left[ {{t_1},{t_2}} \right]^*} =\left[ {\frac{{28{m^2}{T_s}}}{{120{T_s} - 74{m^2}}},{T_s}} \right],
\end{align}
where $m = \frac{d}{{\sqrt {4D} }}$.

% Note that in \eqref{final_t1_t2_solution_absorbing_SID_L_1}, the reason why we just provide an approximate solution for ${\left[ {{t_1},{t_2}} \right]^*}$ is that in all propositions described, the closed-form expression of ${\left[ {{t_1},{t_2}} \right]^*}$ is originally deduced from mSINAR, while the objective function generated from mSINAR is only the approximation of \eqref{objective_AB}.

\emph{Proof:} See Appendix A.

\textbf{\emph{Proposition 2:}} Assuming that the approximation for \eqref{find_interval_AB} with $L>1$ is exact, the solution of \eqref{find_interval_AB} with $L > 1$ and $0 < {\cal Q} < \hat {\cal Q}$ can be expressed as
\begin{align}\label{final_t1_t2_solution_absorbing_SID_L_more_than_1}
{\left[ {{t_1},{t_2}} \right]^*} = \left\{ {\begin{array}{*{20}{c}}
   {\left[ {\frac{{28{m^2}{T_s}}}{{120{T_s} - 28{T_s}\ln {\cal I} - 74{m^2}}},{T_s}} \right],~{\rm{if}}~\sum\limits_{k = 1}^L {\frac{1}{{\sqrt {{{\left( {1 + k} \right)}^3}} }}} \exp \left( {\frac{k}{{\left( {1 + k} \right)}}\frac{{{m^2}}}{{{T_s}}}} \right) \le 1}
   \\ \vspace{-0.1cm}\\
   {\left\{ \begin{array}{l}
 \left[ {\frac{{28{m^2}{T_s}}}{{120{T_s} - 28{T_s}\ln {\cal I} - 74{m^2}}}, \frac{{ - 3 + \left( {\sqrt[3]{{{s_1}}} + \sqrt[3]{{{s_2}}}} \right)}}{3}{T_s}} \right],~{\rm{if}}~{\Delta _1} < 0
 \\ \vspace{-0.5cm}\\
 \left[ {\frac{{28{m^2}{T_s}}}{{120{T_s} - 28{T_s}\ln {\cal I} - 74{m^2}}},\frac{{ - \gamma  + \sqrt {{\Delta _2}} }}{6{\ln {\mathcal {V}} }}{T_s}} \right],~{\rm{if}}~{\Delta _1} \ge 0, {\Delta _2} \ge 0 ,
 \\ \vspace{-0.5cm}\\
 \left[ {\frac{{28{m^2}{T_s}}}{{120{T_s} - 28{T_s}\ln {\cal I} - 74{m^2}}},-\frac{{  \gamma }}{6{\ln {\mathcal {V}} }}{T_s}} \right],~{\rm{if}}~{\Delta _1} \ge 0,{\Delta _2} < 0 \\
 \end{array} \right.}~{\rm others}
 \\ \vspace{-0.5cm}\\
\end{array}} \right.
\end{align}
where the above symbols, such as $\mathcal I$ and $\mathcal V$, have defined in Appendix B.

\emph{Proof:} See Appendix B.

% Note that the results in \textbf{\emph{Propositions 1$\&$2}} give an approximate but good solution for the optimal ${\left[ {{t_1},{t_2}} \right]}$ when the absorbing receiver is considered. Due to the complex objective function, we have to take some approximations to obtain a closed-form ${\left[ {{t_1},{t_2}} \right]}^*$. Importantly, the above solutions have been validated in Section IV-B.

Note that the results in \textbf{\emph{Propositions 1$\&$2}} give an approximate but good solution for the optimal ${\left[ {{t_1},{t_2}} \right]}$, which has been validated via the numerical simulation in Section IV-B.

% According to \textbf{\emph{Propositions 1-2}}, ${\left[ {{t_1},{t_2}} \right]^*}$ derived from the combination of mSINAR and SID can be expressed as
% \begin{align}\label{mSINAR_SID_combine}
% {\left[ {{t_1},{t_2}} \right]^*} \approx \left[ {\frac{{28{m^2}{T_s}}}{{120{T_s} - 28{T_s}\ln {\mathcal I} - 74{m^2}}},\min \left[ {f\left( {{t_2}} \right)h\left( {{t_2}} \right) - \int_{t_1^*}^{{t_2}} {h\left( t \right)dtf'\left( {{t_2}} \right)}  = 0,{T_s}} \right]} \right].
% \end{align}
% Although \eqref{mSINAR_SID_combine} only gives a semi-closed solution for ${\left[ {{t_1},{t_2}} \right]^*}$, it can provide a reference to measure the performance of ${\left[ {{t_1},{t_2}} \right]^*}$ solved by SID. Moreover, it is worth noting that
% both the semi-closed form of ${\left[ {{t_1},{t_2}} \right]^*}$ in \eqref{mSINAR_SID_combine} and

\subsubsection{${\cal Q} \ge {\hat {\cal Q}}$}

In this subsection, we will derive an optimal ${\left[ {{t_1},{t_2}} \right]}$ when ${\cal Q} \ge {\hat {\cal Q}}$. However, how to determine the exact value of ${\hat {\cal Q}}$ is a key problem. In the previous subsection, ${\left[ {{t_1},{t_2}} \right]}^*$ with $0 < {\cal Q} < \hat {\cal Q}$ has been obtained and it is independent of $\cal Q$. Hence, we assume that ${\left[ {{t_1},{t_2}} \right]}^*$ is also fit for ${\hat {\cal Q}}$, and then ${\hat {\cal Q}}$ can be determined directly. Next, we focus on how to calculate ${\left[ {{t_1},{t_2}} \right]}^*$. Constrained by the computational complexity of \eqref{mSINAR_objective}, we convert mSINAR to mSID defined earlier to simplify the optimization procedure.
% First, we define mSID as the difference between the number of desired molecules and undesired molecules contributing by the ISI signal and noise, mathematically expressed as
% \begin{align}\label{mSID_definition}
% {\textrm{mSID}} = F_{ab}^0 - \sum\limits_{k = 1}^L {F_{ab}^k}  - \sqrt {\frac{2}{\cal Q}} \sum\limits_{k = 0}^L {\sqrt {F_{ab}^k\left( {1 - F_{ab}^k} \right)} } .
% \end{align}
When ${\cal Q} \ge {\hat {\cal Q}}$, \eqref{mSINAR_objective} can be transformed into
\begin{align}\label{mSID_objective}
 {\left[ {{t_1},{t_2}} \right]^*} &\approx \mathop {\arg \max }\limits_{0 \le {t_1},{t_2} \le {T_s}} F_{ab}^0 - \sum\limits_{k = 1}^L {F_{ab}^k}  - \sqrt{\frac{2}{\cal Q}}\sum\limits_{k = 0}^L {\sqrt {{{F_{ab}^k}}{}} }  \nonumber\\
  &= \mathop {\arg \max }\limits_{0 \le {t_1},{t_2} \le {T_s}}\int_{{t_1}}^{{t_2}} {\left[ {h\left( t \right) - \sum\limits_{k = 1}^L {h\left( {k{T_s} + t} \right)} } \right]dt}  - \sqrt{\frac{2}{\cal Q}}\sum\limits_{k = 0}^L {\sqrt {{{F_{ab}^k}\left( {1 - F_{ab}^k} \right)}{}} }.
\end{align}
It is considered that the expected signal plays a leading role in \eqref{mSID_objective}, since the true maximum of mSINAR is greater than 1 for ${\cal Q} \ge {\hat {\cal Q}}$. Followed by the above fact, we assume that mSID can have a certain tolerance threshold for the change of the standard variance of the noise. Therefore, for ${\sqrt {F_{ab}^k} }\left( {1 - F_{ab}^k} \right)$, we have
\begin{align}
 \sqrt {F_{ab}^0\left( {1 - F_{ab}^0} \right)}  \approx {\alpha _1}F_{ab}^0,~~\sqrt {F_{ab}^k\left( {1 - F_{ab}^k} \right)} \approx {\alpha _2}F_{ab}^k,
 \end{align}
To ensure that the above approximations are feasible for any possible ${\left[ {{t_1},{t_2}} \right]}$, we define ${\alpha _1}$ and ${\alpha _2}$ as the minimum of all possible $\frac{{\sqrt {F_{ab}^0\left( {1 - F_{ab}^0} \right)} }}{{F_{ab}^0}}$ and $\frac{{\sqrt {F_{ab}^k\left( {1 - F_{ab}^k} \right)} }}{{F_{ab}^k}}$ with $k = 1, \cdots L$, respectively,~i.e.,
\begin{align}
{\alpha _1} = \sqrt {\frac{{1 - \int_0^{{T_s}} {h\left( t \right)} dt}}{{\int_0^{{T_s}} {h\left( t \right)} dt}}} , ~{\alpha _2} = \sqrt {\frac{{1 - \int_0^{{T_s}} {h\left( {t + {T_s}} \right)} dt}}{{\int_0^{{T_s}} {h\left( {t + {T_s}} \right)} dt}}} .
\end{align}
At this point, \eqref{mSID_objective} can be simplified as
\begin{align}\label{mSID_objective_simplified}
{\left[ {{t_1},{t_2}} \right]^*} \approx \mathop {\arg \max }\limits_{0 \le {t_1},{t_2} \le {T_s}} \int_{{t_1}}^{{t_2}} {\left[ {h\left( t \right) - {\cal G}\left( {\cal Q} \right)\sum\limits_{k = 1}^L {h\left( {k{T_s} + t} \right)} } \right]dt} ,
\end{align}
where ${\cal G}\left( {\cal Q} \right) = {\frac{{\sqrt {\cal Q}  + \sqrt{2}{\alpha _2}}}{{\sqrt {\cal Q}  - \sqrt{2}{\alpha _1}}}}$. As can be observed from \eqref{mSINAR_objective}, we have ${\cal Q} = {\hat {\cal Q}}$ when calculating ${\left[ {{t_1},{t_2}} \right]^*}$ for ${\cal Q} \ge {\hat {\cal Q}}$. Certainly, ${\left[ {{t_1},{t_2}} \right]}^*$ with ${\cal Q} = {\hat {\cal Q}}$ is suitable for all possible $\cal Q$ in the range of $\left[ {\hat {\cal Q},\infty } \right)$. However, if ${\cal Q} = {\hat {\cal Q}}$ is true for mSID, \eqref{mSID_objective_simplified} can be rewritten~as
\begin{align}\label{special_mSID_objective_simplified}
\mathop {\arg \max }\limits_{0 \le {t_1},{t_2} \le {T_s}} \int_{{t_1}}^{{t_2}} {\left[ {h\left( t \right) - {\cal G}\left( {\cal {\hat Q}} \right)\sum\limits_{k = 1}^L {h\left( {k{T_s} + t} \right)} } \right]dt} = 0.
\end{align}
Here, for the same reason clarified earlier, the solution of \eqref{special_mSID_objective_simplified} is either $t_1^* = t_2^* = {t_{\max }}$ or no solution. To construct a function where ${\left[ {{t_1},{t_2}} \right]^*}$ can be approximately solved from \eqref{mSID_objective_simplified}, we decide to slightly enlarge the maximum of mSINAR to obtain a new ${ {\cal Q}}$ and then make \eqref{mSID_objective_simplified} greater than 0. Therefore, we define ${\cal Q} =
{\cal G}\left( {\cal {\hat Q}} \right){\hat {\cal Q}}$. Here, \eqref{mSID_objective_simplified} is equivalent to collecting all of intervals satisfying $ h{\left( t \right) - {\cal G}\left( {\cal Q} \right)\sum\limits_{k = 1}^L {h\left( {k{T_s} + t} \right)}} \ge 0$, and we have
\begin{align}\label{find_interval_AB_mSID}
h{\left( t \right) - {\cal G}\left( {\cal Q} \right)\sum\limits_{k = 1}^L {h\left( {k{T_s} + t} \right)} } = 0 , i=1,2.
\end{align}
For clarity, the ${\left[ {{t_1},{t_2}} \right]^*}$ obtained from \eqref{find_interval_AB_mSID} has been described in the following proposition.

\textbf{\emph{Proposition 3:}} Assuming that the approximation for \eqref{find_interval_AB_mSID} for all considered $L$ is exact, the solution of \eqref{find_interval_AB_mSID} with $L > 1$ and $\cal Q \ge {\hat {\cal Q}}$ can be written as
\begin{align}\label{Final_optimal_interval_AB}
{\left[ {{t_1},{t_2}} \right]^*} = \left[ {\frac{{28{m^2}{T_s}}}{{120{T_s} - 28{T_s}\ln {\cal G}\left( {\cal Q} \right) - 74{m^2}}},{T_s}} \right].
\end{align}
As for $L>1$ and $\cal Q \ge {\hat {\cal Q}}$, the solution of \eqref{find_interval_AB_mSID} can be expressed as
\begin{align}\label{final_t1_t2_solution_absorbing}
{\left[ {{t_1},{t_2}} \right]^*} = \left\{ {\begin{array}{*{20}{c}}
   {\left[ {\frac{{28{m^2}{T_s}}}{{120{T_s} - 28{T_s}\ln {\cal I}{\cal G}\left( {\cal Q} \right) - 74{m^2}}},{T_s}} \right],~{\rm{if}}~{\cal G}\left( {\cal Q} \right)\sum\limits_{k = 1}^L {\frac{1}{{\sqrt {{{\left( {1 + k} \right)}^3}} }}} \exp \left( {\frac{k}{{\left( {1 + k} \right)}}\frac{{{m^2}}}{{{T_s}}}} \right) \le 1}
   \\ \vspace{-0.1cm}\\
   {\left\{ \begin{array}{l}
 \left[ {\frac{{28{m^2}{T_s}}}{{120{T_s} - 28{T_s}\ln {\cal I}{\cal G}\left( {\cal Q} \right) - 74{m^2}}}, \frac{{ - 3 + \left( {\sqrt[3]{{{s_1}}} + \sqrt[3]{{{s_2}}}} \right)}}{3}{T_s}} \right],~{\rm{if}}~{\Delta _1} < 0
 \\ \vspace{-0.5cm}\\
 \left[ {\frac{{28{m^2}{T_s}}}{{120{T_s} - 28{T_s}\ln {\cal I}{\cal G}\left( {\cal Q} \right) - 74{m^2}}},\frac{{ - \gamma  + \sqrt {{\Delta _2}} }}{6{\ln {\mathcal {V}}{\cal G}\left( {\cal Q} \right) }}{T_s}} \right],~{\rm{if}}~{\Delta _1} \ge 0, {\Delta _2} \ge 0 ,
 \\ \vspace{-0.5cm}\\
 \left[ {\frac{{28{m^2}{T_s}}}{{120{T_s} - 28{T_s}\ln {\cal I}{\cal G}\left( {\cal Q} \right) - 74{m^2}}},-\frac{{  \gamma }}{6{\ln {\mathcal {V}}{\cal G}\left( {\cal Q} \right) }}{T_s}} \right],~{\rm{if}}~{\Delta _1} \ge 0,{\Delta _2} < 0 \\
 \end{array} \right.}~{\rm others}
 \\ \vspace{-0.5cm}\\
\end{array}} \right.
\end{align}
\emph{Proof:} Since the calculation of \eqref{find_interval_AB_mSID} is  similar to \eqref{find_interval_AB} in addition to the introduction of ${\cal G}\left( {\cal Q} \right)$, please see Appendix B about the detailed derivation of ${\left[ {{t_1},{t_2}} \right]^*} $. Moreover, it is worth noting that $\cal I$ and $\cal V$ are replaced by ${\cal I}{\cal G}\left( {\cal Q} \right)$ and ${\cal V}{\cal G}\left( {\cal Q} \right)$, respectively, in $\gamma$, ${\Delta _1}$, and ${\Delta _2}$. Furthermore, $\hat t_1^*$ in \eqref{L>1_AB_simple_t1} and $\hat t_2^*$ in \eqref{ratio_AB_t2} are updated as $\hat t_1^* = {t_1^*}$ and $\hat t_2^* = \frac{1}{2}\left({t_2^* + t_{\textrm{max}}}\right)$, where $t_1^*$ and $t_1^*$ are generated from the derived ${\left[ {{t_1},{t_2}} \right]^*}$ with $0<\cal Q < {\hat {\cal Q}}$.

% Similar to \eqref{mSINAR_SID_combine}, when $\cal Q \ge {\hat {\cal Q}}$, we also can obtain ${\left[ {{t_1},{t_2}} \right]^*}$ derived from the combination of mSINAR and mSID, mathematically expressed as
% \begin{align}\label{mSINAR_mSID_combine}
% \hspace{-0.25cm}{\left[ {{t_1},{t_2}} \right]^*} \hspace{-0.05cm}=\hspace{-0.05cm} \left[\hspace{-0.09cm} {\frac{{28{m^2}{T_s}}}{{120{T_s} \hspace{-0.05cm}-\hspace{-0.05cm} 28{T_s}\ln {\cal I}{\cal G}\left( {\cal Q} \right) \hspace{-0.05cm}-\hspace{-0.05cm} 74{m^2}}},\min\hspace{-0.05cm}\left[ {f\left( {{t_2}} \right)h\left( {{t_2}} \right) \hspace{-0.05cm}-\hspace{-0.08cm} \int_{t_1^*}^{{t_2}} {h\left( t \right)} dtf'\left( {{t_2}} \right) \hspace{-0.05cm}=\hspace{-0.05cm} 0,{T_s}} \right]}\hspace{-0.05cm}\right].
% \end{align}

\subsection{Detection Interval Optimization for Passive Receiver}

In this subsection, we focus on the optimal detection interval for the passive receiver.
%Further, ${\xi _{opt}}$ is the average optimal detection threshold and it can be obtained via the similar method of \eqref{optimal_threshold}.
%Next, we focus on how to compute the optimal sampling interval ${\left[ {{n_1},{n_2}} \right]^*}$.
According to the description on the 3D absorbing receiver, the objective function can be written~as
\begin{align}\label{detection_interval_passive}
{\left[ {{n_1},{n_2}} \right]^*} &= \mathop {\arg \min }\limits_{0 \le {n_1},{n_2} \le N} {P_e} \nonumber\\
&\approx \left\{ \begin{array}{l}
 \mathop {\arg \max }\limits_{0 \le {n_1},{n_2} \le N} \frac{{\sum\limits_{n = {n_1}}^{{n_2}} {{p_{n,0}}} }}{{\sum\limits_{k = 1}^L {\sum\limits_{n = {n_1}}^{{n_2}} {{p_{n,k}}} }  + \sum\limits_{k = 0}^L {\sqrt {\frac{2}{{\cal Q}}\sum\limits_{n = {n_1}}^{{n_2}} {{p_{n,k}}} } } }},~{\textrm{if}}~0 < {\cal Q} < \hat {\cal Q} \\
 \mathop {\arg \max }\limits_{0 \le {n_1},{n_2} \le N} \frac{{\sum\limits_{n = {n_1}}^{{n_2}} {{p_{n,0}}} }}{{\sum\limits_{k = 1}^L {\sum\limits_{n = {n_1}}^{{n_2}} {{p_{n,k}}} }  + \sum\limits_{k = 0}^L {\sqrt {\frac{2}{{\hat {\cal Q}}}\sum\limits_{n = {n_1}}^{{n_2}} {{p_{n,k}}} } } }},~{\textrm{if}}~{\cal Q} \ge \hat {\cal Q}
 \end{array} \right.,
\end{align}
where $\hat {\cal Q} = \left\lceil {2{{\left( {\frac{{\sum\limits_{k = 0}^L {\sqrt {\sum\limits_{n = {n_1}}^{{n_2}} {{p_{n,k}}} } } }}{{\sum\limits_{n = {n_1}}^{{n_2}} {{p_{n,0}}}  - \sum\limits_{k = 1}^L {\sum\limits_{n = {n_1}}^{{n_2}} {{p_{n,k}}} } }}} \right)}^2}} \right\rceil $,
% \begin{align}
% \hat {\cal Q} = \left\lceil {2{{\left( {\frac{{\sum\limits_{k = 0}^L {\sqrt {\sum\limits_{n = {n_1}}^{{n_2}} {{p_{n,k}}} } } }}{{\sum\limits_{n = {n_1}}^{{n_2}} {{p_{n,0}}}  - \sum\limits_{k = 1}^L {\sum\limits_{n = {n_1}}^{{n_2}} {{p_{n,k}}} } }}} \right)}^2}} \right\rceil \nonumber.
% \end{align}
calculated from \eqref{detection_interval_passive} when assuming mSINAR = 1, where ${\left[ {{n_1},{n_2}} \right]^*}$ is also assumed to be used. Subsequently, we will show the detailed derivation for the optimal ${\left[ {{n_1},{n_2}} \right]}$.

\subsubsection{$0 < {\cal Q} < \hat {\cal Q}$}  In this subsection, we use the SID to obtain an approximate ${\left[ {{n_1},{n_2}} \right]^*}$ with $0 < {\cal Q} < \hat {\cal Q}$, when there is always a $\left[ {{n_1},{n_2}} \right]$ to make
\begin{align}
\sum\limits_{n = {n_1}}^{{n_2}} {{p_{n,0}}}  - \sum\limits_{k = 1}^L {\sum\limits_{n = {n_1}}^{{n_2}} {{p_{n,k}}} }  > 0 \nonumber
\end{align}
hold. Here, the corresponding objective function can be re-written as
\begin{align}\label{detection_interval_passive_mSID}
{\left[ {{n_1},{n_2}} \right]^*} \approx \mathop {\arg \max }\limits_{0 \le {n_1},{n_2} \le N} \sum\limits_{n = {n_1}}^{{n_2}} {{p_{n,0}} - \sum\limits_{k = 1}^L {\sum\limits_{n = {n_1}}^{{n_2}} {{p_{n,k}}} } } .
\end{align}
It is clear from \eqref{detection_interval_passive_mSID} that the condition for maximizing $\sum\limits_{n = {n_1}}^{{n_2}} {{p_{n,0}} - \sum\limits_{k = 1}^L {\sum\limits_{n = {n_1}}^{{n_2}} {{p_{n,k}}} } } $ is to ensure $\left({p_{n,0}} - \sum\limits_{k = 1}^L {{p_{n,k}}} \right)>0$, for $n \in {\left[ {{n_1},{n_2}} \right]^*}$. This means if
\begin{align}\label{find_interval_passive}
{p_{n_j^*,0}} = \sum\limits_{k = 1}^L {{p_{n_j^*,k}}} ,~j=1,2
\end{align}
can be solved, we can obtain $ {\left[ {{n_1},{n_2}} \right]^*}$. For clarity, \eqref{find_interval_passive} can be re-arranged as
\begin{align}\label{find_interval_passive_simple}
\sum\limits_{k = 1}^L \frac{1}{{{{\left( {{n_j^*}{t_s} + k{T_s}} \right)}^{3/2}}}} \exp \left( { - \frac{{{{(d + r)}^2}}}{{4D\left( {{n_j^*}{t_s} + k{T_s}} \right)}}} \right)-\frac{1}{{{{\left( {{n_j^*}{t_s}} \right)}^{3/2}}}}\exp \left( { - \frac{{{{(d + r)}^2}}}{{4D{n_j^*}{t_s}}}} \right) =0,
\end{align}
where $j=1,2$. Constrained by the computational complexity, we also solve \eqref{find_interval_passive_simple} according to the ISI length $L$.

\textbf{\emph{Proposition 4:}} Assuming that the approximation for \eqref{find_interval_passive_simple} for all considered $L$ is exact, the solution of \eqref{find_interval_passive_simple} with $L = 1$ and $0 < {\cal Q} < \hat {\cal Q}$ can be solved as
\begin{align}\label{Final_optimal_interval_Passive_L_1}
{\left[ {{n_1},{n_2}} \right]^*} = \left[ {\left\lceil {\frac{{28{{\hat m}^2}{T_s}}}{{\left( {120{T_s} - 74{{\hat m}^2}} \right){t_s}}}} \right\rceil,N} \right].
\end{align}
When $L>1$ and $0 < {\cal Q} < \hat {\cal Q}$, the solution of \eqref{find_interval_passive_simple} can be stated as
\begin{align}\label{optimal_n1_n2_SID}
{\left[ {{n_1},{n_2}} \right]^*} = \left\{ \begin{array}{l}
 \left[ {\left\lceil {\frac{{28{{\hat m}^2}{T_s}}}{{\left( {120{T_s} - 28{T_s}\ln {\cal{ W}} - 74{{\hat m}^2}} \right){t_s}}}} \right\rceil ,N} \right],~{\rm{if}}~\sum\limits_{k = 1}^L {\frac{1}{{\sqrt {{{\left( {1 + k} \right)}^3}} }}} \exp \left( {\frac{k}{{\left( {1 + k} \right)}}\frac{{{{\hat m}^2}}}{{{T_s}}}} \right) \le 1 \\
  \vspace{-0.3cm} \\
 \left[ {\left\lceil {\frac{{28{{\hat m}^2}{T_s}}}{{\left( {120{T_s} - 28{T_s}\ln {\cal{W}} - 74{{\hat m}^2}} \right){t_s}}}} \right\rceil ,\left\lceil {\frac{{ - 3 + \left( {\sqrt[3]{{{{\hat s}_1}}} + \sqrt[3]{{{{\hat s}_2}}}} \right)}}{3}N} \right\rceil } \right],~{\rm{if}}~{{\hat \Delta }_1} < 0 \\
  \vspace{-0.3cm} \\
 \left[ {\left\lceil {\frac{{28{{\hat m}^2}{T_s}}}{{\left( {120{T_s} - 28{T_s}\ln {\cal{W}} - 74{{\hat m}^2}} \right){t_s}}}} \right\rceil ,\left\lceil {\frac{{ - \hat \gamma  + \sqrt {{{\hat \Delta }_2}} }}{{6\ln \cal{A}}}N} \right\rceil } \right],~{\rm{if}}~{{\hat \Delta }_1} \ge 0,{{\hat \Delta }_2} \ge 0 \\
  \vspace{-0.3cm} \\
 \left[ {\left\lceil {\frac{{28{{\hat m}^2}{T_s}}}{{\left( {120{T_s} - 28{T_s}\ln {\cal{W}} - 74{{\hat m}^2}} \right){t_s}}}} \right\rceil ,\left\lceil {\frac{{ - \hat \gamma }}{{6\ln \cal{A}}}N} \right\rceil } \right],~{\rm{if}}~{{\hat \Delta }_1} \ge 0,{{\hat \Delta }_2} < 0 \\
 \end{array} \right..
\end{align}

\emph{Proof:} See Appendix C.

\subsubsection{${\cal Q} \ge \hat {\cal Q}$}

In this subsection, we will derive an optimal ${\left[ {{n_1},{n_2}} \right]}$, which is suitable for all possible $\cal Q$ in the range of $\left[ {\hat {\cal Q},\infty } \right)$. Similarly, we use ${\left[ {{n_1},{n_2}} \right]^*}$ derived when $0 < {\cal Q} < \hat {\cal Q}$ to determine ${\hat {\cal Q}}$. Then, by using the mSID, we can transform the objective function in \eqref{detection_interval_passive} into
\begin{align}\label{detection_interval_passive_mSID_Q_larger}
{\left[ {{n_1},{n_2}} \right]^*} \approx \mathop {\arg \max }\limits_{0 \le {n_1},{n_2} \le N} \left\{\sum\limits_{n = {n_1}}^{{n_2}} {{p_{n,0}} - \hat {\cal G}\left( {\cal Q} \right)\sum\limits_{k = 1}^L {\sum\limits_{n = {n_1}}^{{n_2}} {{p_{n,k}}} } }\right\} ,
\end{align}
where $\hat {\cal G}\left( {\cal Q} \right) = \frac{{\left( {\sqrt {\cal Q}  + \sqrt 2 {{\hat \alpha }_2}} \right)}}{{\left( {\sqrt {\cal Q}  - \sqrt 2 {{\hat \alpha }_1}} \right)}}$, and
\begin{align}
{{\hat \alpha }_1} \approx \frac{1}{{\sqrt {\sum\limits_{n = 0}^N {{p_{n,0}}} } }},~{{\hat \alpha }_2} \approx \frac{1}{{\sqrt {\sum\limits_{n = 0}^N {{p_{n,{\rm{1}}}}} } }}.  \nonumber
\end{align}
In order to guarantee the feasibility of mSID, we also set ${\cal Q} =\hat {\cal G}\left( \hat {\cal Q} \right)\hat {\cal Q}$ to solve \eqref{detection_interval_passive_mSID_Q_larger}.
It is clear from \eqref{detection_interval_passive_mSID_Q_larger} that the condition for maximizing $\sum\limits_{n = {n_1}}^{{n_2}} {{p_{n,0}} - \hat {\cal G}\left( {\cal Q} \right)\sum\limits_{k = 1}^L {\sum\limits_{n = {n_1}}^{{n_2}} {{p_{n,k}}} } } $ is to ensure $\left({p_{n,0}} - \hat {\cal G}\left( {\cal Q} \right)\sum\limits_{k = 1}^L {{p_{n,k}}} \right)>0$, for $n \in {\left[ {{n_1},{n_2}} \right]^*}$. Therefore, \eqref{detection_interval_passive_mSID_Q_larger} can be re-arranged as
\begin{align}\label{find_interval_passive_simple_Q_greater_than}
\hspace{-0.3cm}\hat {\cal G}\left( {\cal Q} \right)\sum\limits_{k = 1}^L \frac{1}{{{{\left( {{n_j^*}{t_s} + k{T_s}} \right)}^{3/2}}}} \exp \left( { - \frac{{{{(d + r)}^2}}}{{4D\left( {{n_j^*}{t_s} + k{T_s}} \right)}}} \right)-\frac{1}{{{{\left( {{n_j^*}{t_s}} \right)}^{3/2}}}}\exp \left( { - \frac{{{{(d + r)}^2}}}{{4D{n_j^*}{t_s}}}} \right) =0,
\end{align}
where $j=1,2$. The calculation of ${\left[ {{n_1},{n_2}} \right]^*}$ in \eqref{detection_interval_passive_mSID_Q_larger} is almost the same to \eqref{detection_interval_passive_mSID}, therefore, ${\left[ {{n_1},{n_2}} \right]^*}$ can be shown in \eqref{Final_optimal_interval_Passive_L_1} and \eqref{optimal_n1_n2_SID} with the introduction of $\hat {\cal G}\left( {\cal Q} \right)$. Besides, $\hat {n_1^*}$ in \eqref{ratio_passive_n1} and $\hat {n_2^*}$ in \eqref{ratio_passive_n2} are replaced by ${n_1^*}$ and $\frac{{n_2^*{t_s} + {t_{\max }}}}{{2{t_s}}}$, respectively, where ${n_1^*}$ and ${n_2^*}$ have been derived when $0<{\cal Q}<\hat {\cal Q}$.

% \emph{{Remark:}} Generally, in traditional communication theory, as the receiver collects more information from the channel, the detection accuracy gradually improves. However, it can be seen from the above analysis that we have ${\left[ {{t_1},{t_2}} \right]^*}~\textrm{or}~{\left[ {{n_1},{n_2}} \right]^*}{t_s} \in \left[ {0,{T_s}} \right]$, which reflects that the optimal detection interval is only a part of the whole symbol duration. How to explain this in-consistence is still an interesting problem. Below, we take the passive receiver as an example for investigation. The first reason is that in this analysis, we only employ the equal sum detector instead of the optimally weighted sum detector to combine the sample signals. On the other hand, given the correlation between samples, it is impractical to use a maximum likelihood detector. Moreover, the analysis is more challenging since the noise depends on the expected received signal.

\section{Numerical results and analysis} \label{Numerical results and analysis}

\begin{table}[t]
\caption{System Parameters}\label{Table_system-parameters} \centering % ???
\begin{tabular}{lccc}  % {cccc} ???????????left-l,right-r,center-c
%\hline
\toprule[1pt]

  Parameter & Absorbing receiver & Passive receiver \\
\hline
\hline
  Radius of receiver ($r$) & $5 \mu$m & $1 \mu$m \\
  Transmission distance ($d$) & $5 \mu$m &$9 \mu$m \\
  Diffusion coefficient ($D$) & $80\times 10^{-12} \rm{m^2/s}$&$80\times 10^{-12} \rm{m^2/s}$ \\
\bottomrule[1pt]
%\hline
\end{tabular}
\end{table}

In this section, we employ Monte Carlo simulations to demonstrate the BER performance of the proposed mSINAR scheme, and include the shift-$\tau$ method investigated in \cite{The_shift_method_1}, the conventional OOK scheme, and the existing SID and SINAR as benchmarks to measure the system performance, where the optimal detection threshold is assumed to be employed for all schemes. Moreover, in the simulations, we reveal the convergence of the optimal detection interval and also verify the theoretical ${\left[ {{t_1},{t_2}} \right]^*}$ and ${\left[ {{n_1},{n_2}} \right]^*}$ obtained from mSINAR. Finally, the system parameters are listed in Table~\ref{Table_system-parameters} and the sampling interval is set to ${t_s} = \left\lfloor {\frac{{{t_{\max }}}}{6}} \right\rfloor $ for the passive receiver.

\subsection{Convergence of Optimal Detection Interval}

\begin{figure}[t]
    \centering
    \subfigure{
        \includegraphics[width=4.5in]{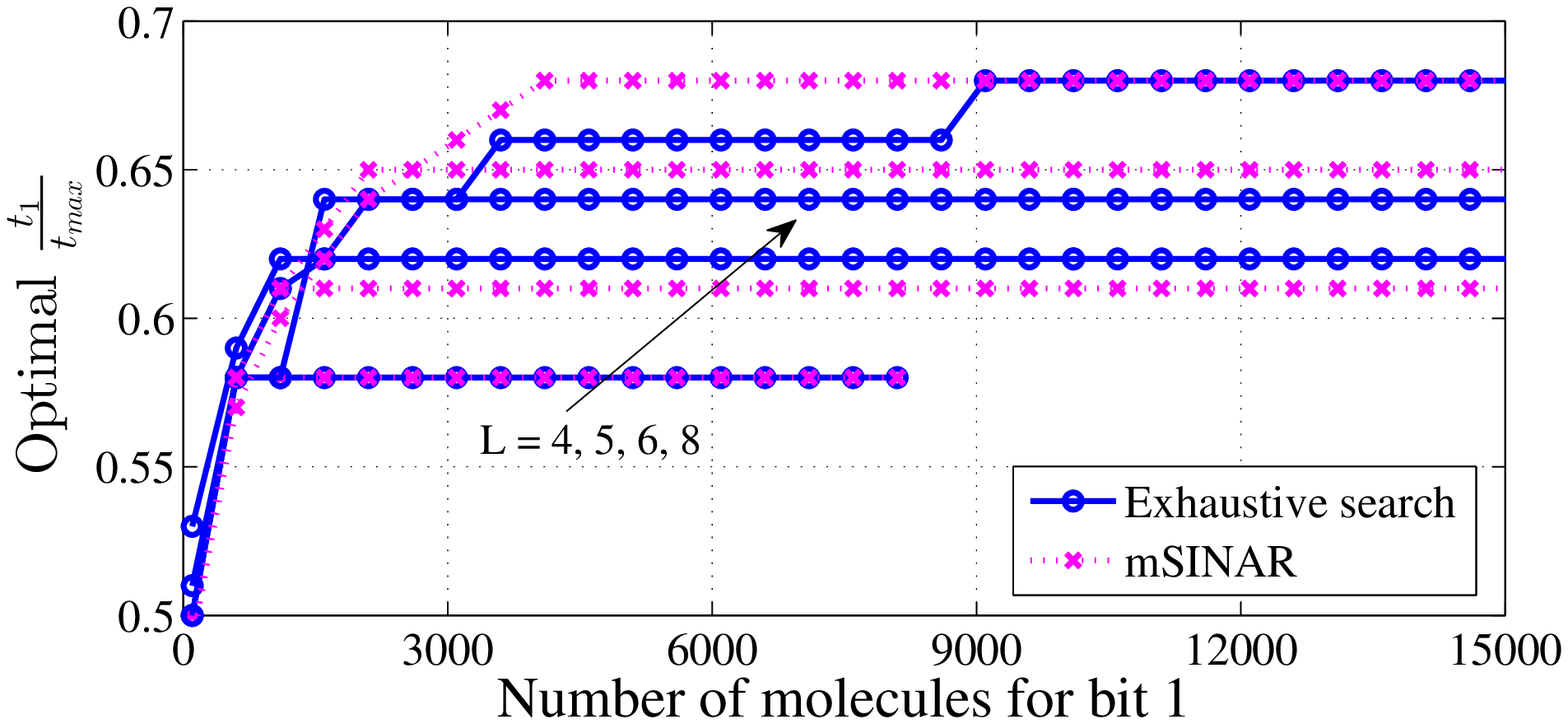}
    \label{Ts_0_2_t1}
    }
    \subfigure{
	\includegraphics[width=4.5in]{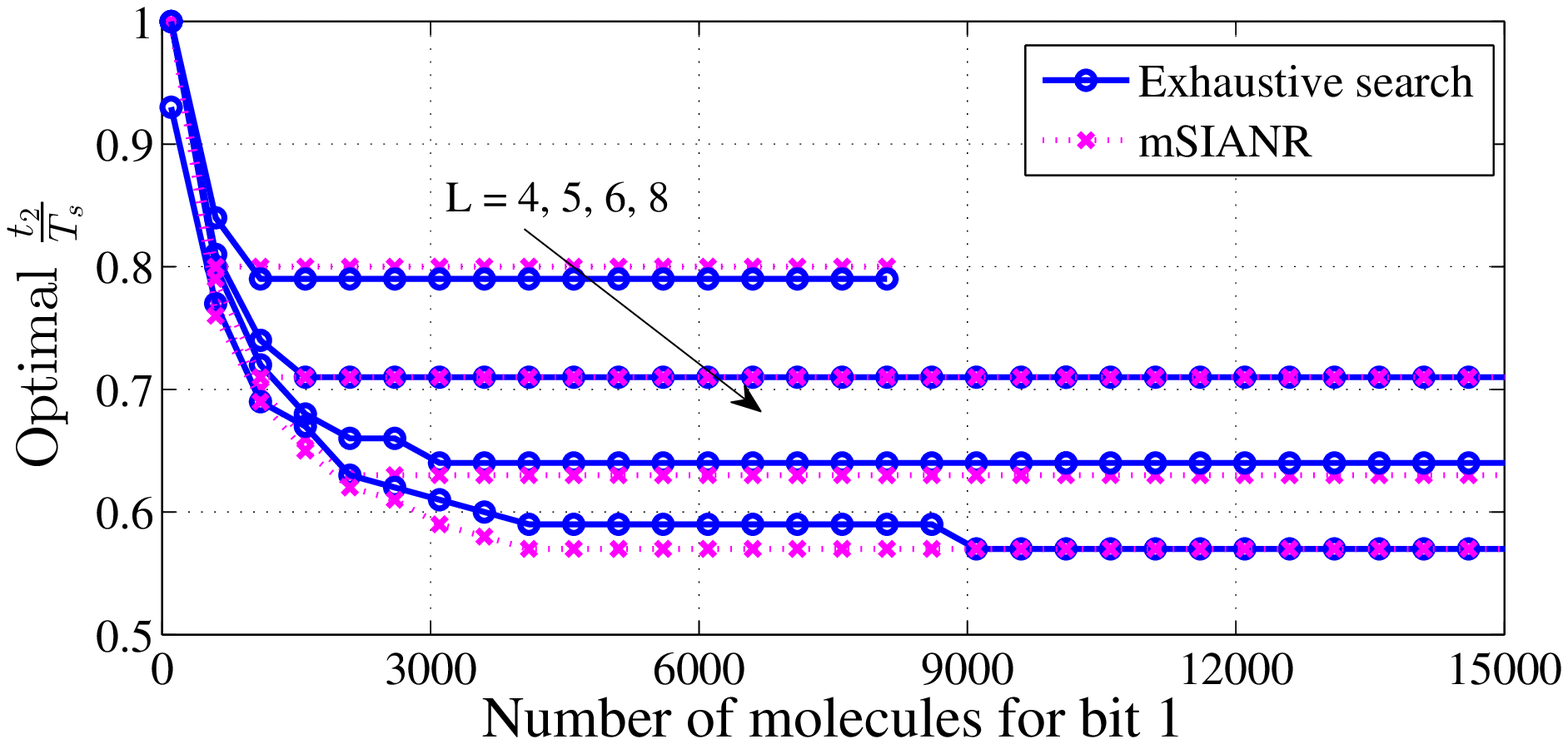}
	\label{Ts_0_2_t2}
    }
    \caption{Convergence of the optimal detection interval ${\left[ {{t_1},{t_2}} \right]^*}$ versus $\cal Q$, where the absorbing receiver with $T_s = 0.2$ and $L = 4,5,6,8$ is considered.}
    \label{Absorbing_receiver_optimal_detection_interval_stability}
\end{figure}

\begin{figure}[t]
    \centering
    \subfigure{
        \includegraphics[width=4.5in]{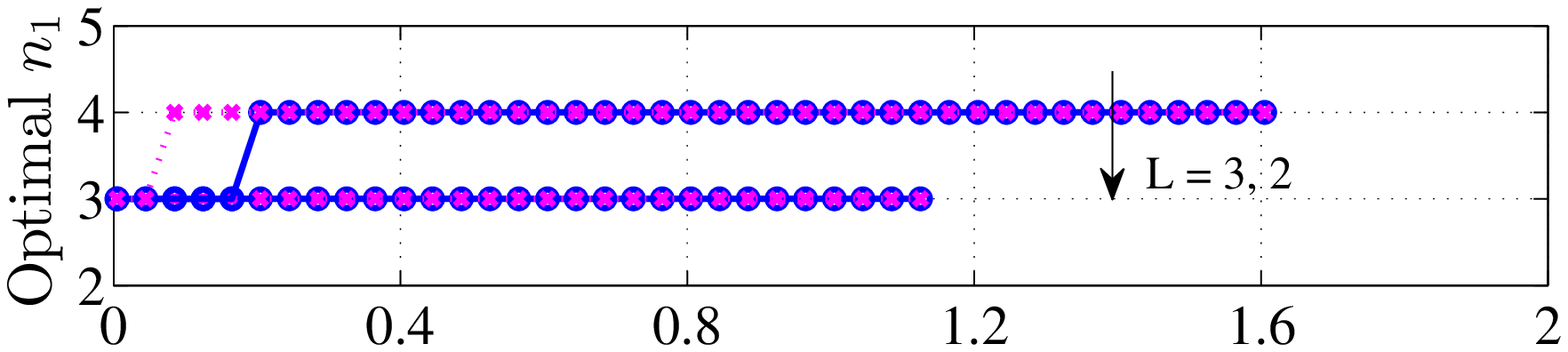}
    \label{Ts_1_n1_2_3}
    }
    \subfigure{
	\includegraphics[width=4.5in]{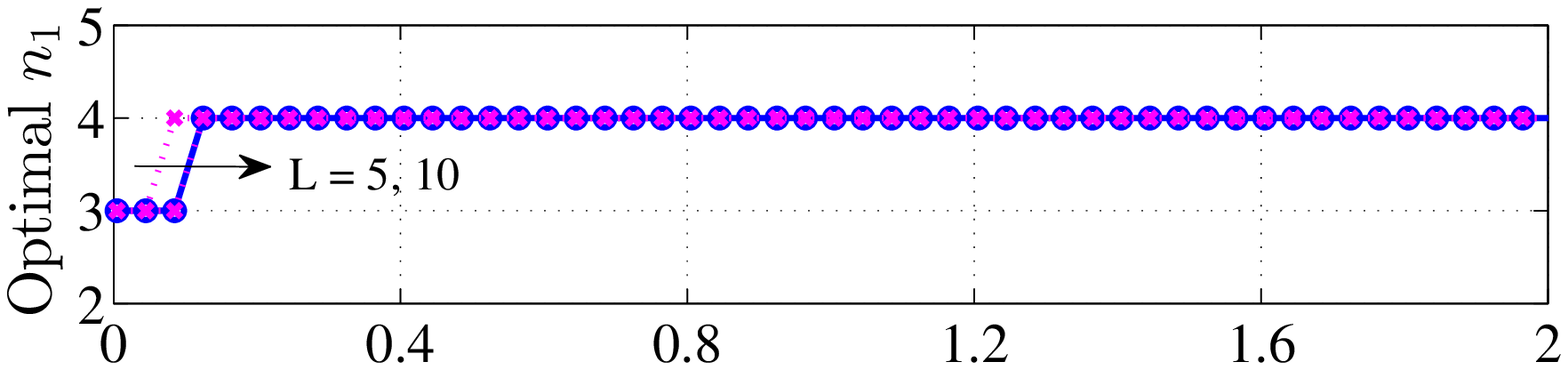}
	\label{Ts_1_n1_5_10}
    }
        \subfigure{
	\includegraphics[width=4.5in]{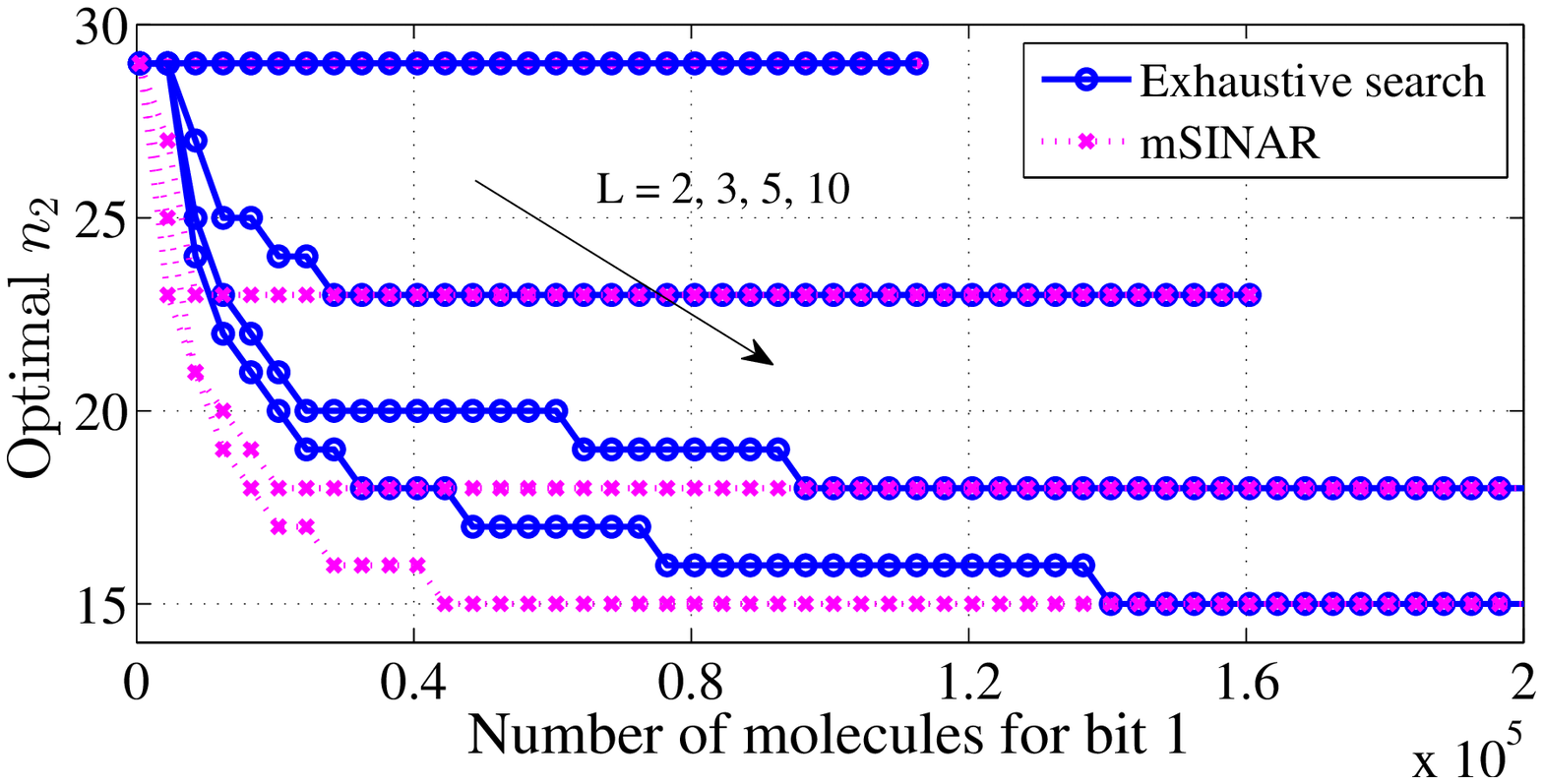}
	\label{Ts_1_n2}
    }
    \caption{Convergence of the optimal detection interval ${\left[ {{n_1},{n_2}} \right]^*}$ versus $\cal Q$, where the passive receiver with $T_s = 1$ and $L = 2,3,5,10$ is considered.}
    \label{Passive_receiver_optimal_detection_interval_stability}
\end{figure}

In this subsection, we investigate the convergence of the optimal detection interval, where the absorbing and passive receivers are considered in Fig.~\ref{Absorbing_receiver_optimal_detection_interval_stability} and Fig.~\ref{Passive_receiver_optimal_detection_interval_stability}, respectively. Besides, we include the optimal ${\left[ {{t_1},{t_2}} \right]}$ and ${\left[ {{n_1},{n_2}} \right]}$ obtained by two different methods for comparison. Taking the absorbing receiver as an example, we refer to ``Exhaustive search" as ${\left[ {{t_1},{t_2}} \right]^*}$ obtained from the exhaustive search for $\arg \min {P_e}$, and ``mSINAR" as ${\left[ {{t_1},{t_2}} \right]^*}$ obtained from \eqref{mSINAR_objective}.

\begin{figure}[t]
    \centering
    \subfigure[absorbing receiver with $T_s = 0.2$]{
        \includegraphics[width=4.5in]{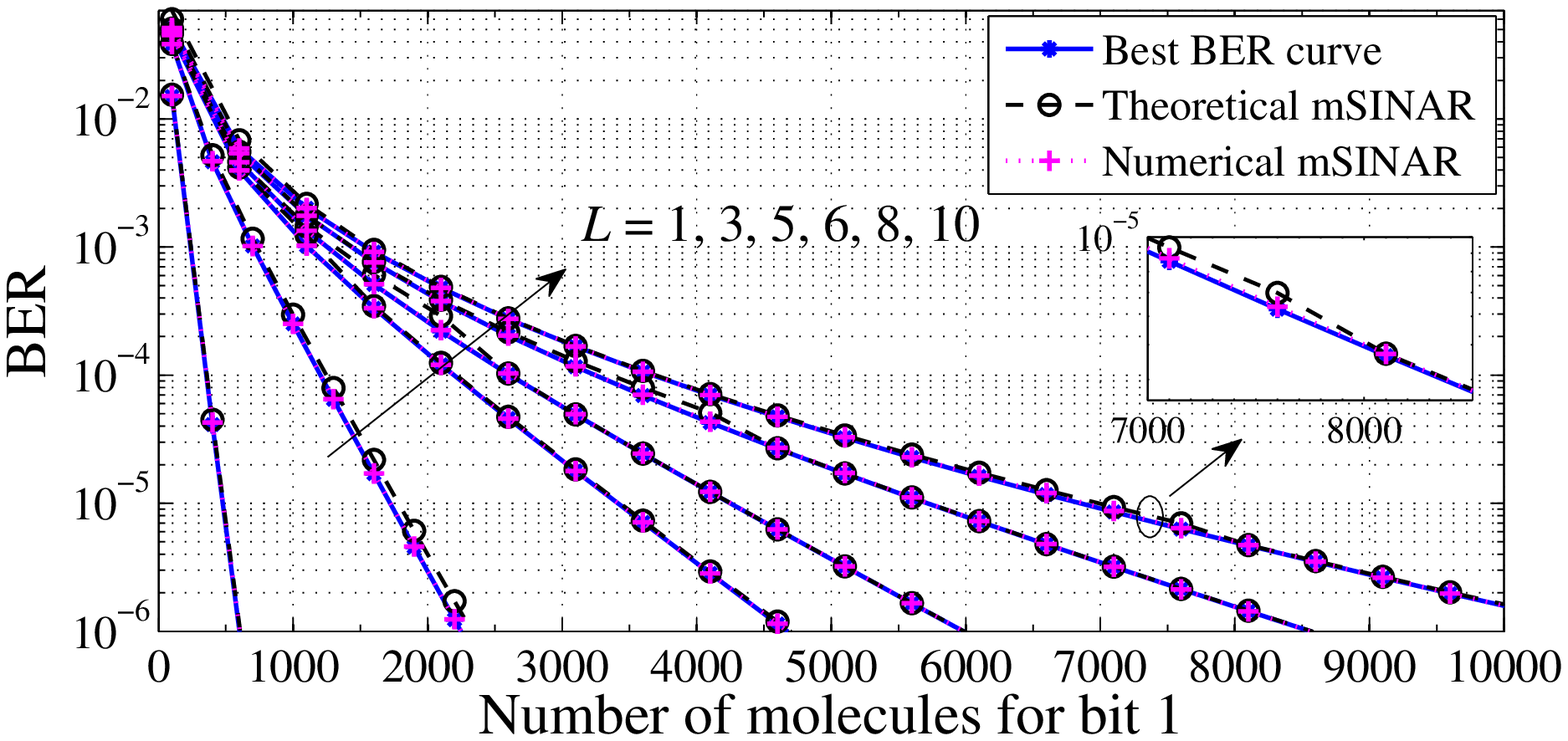}
    \label{Ts_0_2_verification}
    }
        \subfigure[absorbing receiver with $T_s = 0.3$]{
	\includegraphics[width=4.5in]{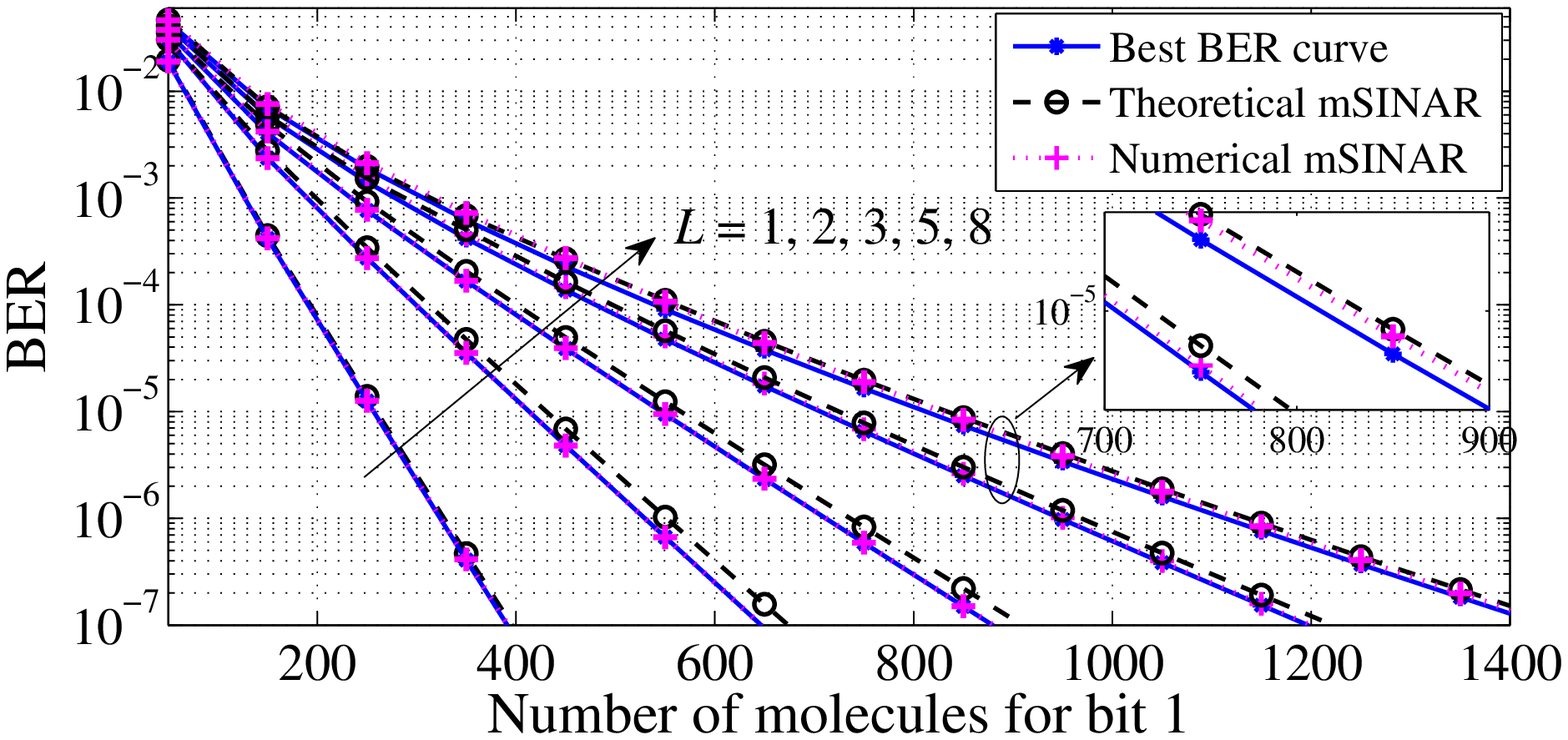}
	\label{Ts_0_3_verification}
	}
    \caption{BER versus $\cal Q$ performance of the proposed mSINAR: numerical verification, where the absorbing receiver with ${T_s} = \left\{ {0.2,0.3} \right\}$ is considered.}
    \label{Absorbing_receiver_verification}
\end{figure}

Fig.~\ref{Absorbing_receiver_optimal_detection_interval_stability} shows the convergence of $t_1^*$ and $t_2^*$, respectively, corresponding to the first and second subplots, where the absorbing receiver with $T_s = 0.2$ and $L = 4,5,6,8$ is considered. It can be seen from these plots that for all cases considered, both $t_1^*$ and $t_2^*$ eventually converge, especially a smaller $L$ will accelerate the convergence. This can be explained as follows:
since the influence of noise is inversely proportional to $\cal Q$, all interference can be approximately regarded as linearly related to the desired signal, when $\cal Q$ is large enough; while once this linear correlation is first achieved, ${\left[ {{t_1},{t_2}} \right]^*}$ remains unchanged for future $\cal Q$. Besides, for a given $\cal Q$, due to the weaker interference, the situation with a smaller $L$ can achieve this linear correlation faster. Moreover, we can observe from Fig.~\ref{Absorbing_receiver_optimal_detection_interval_stability} that the initial value of $t_1^*$ is about $\frac{1}{2}t_{\textrm{max}}$ instead of $t_{\textrm{max}}$; while for $t_2^*$, its original value is $T_s$. This is because, as long as $L$ is greater than 0, the forward ISI (the ISI with $0 \le t \le {t_{\max }}$) always exists and suppresses the expected signal in a short time, no matter how much $\cal Q$ is; while the backward interference (the ISI and noise with ${t_{\max }}< t \le T_s$) is relatively small with a smaller $\cal Q$. Finally, as expected, ${\left[ {{t_1},{t_2}} \right]^*}$ obtained from mSINAR is well matched to that obtained from the exhaustive search, albeit with a slight gap. Fortunately, in Fig.~\ref{Absorbing_receiver_verification}, we can find that this gap has little effect on BER performance.

\begin{figure}[t]
    \centering
    \subfigure[absorbing receiver with $T_s = 1$]{
        \includegraphics[width=4.5in]{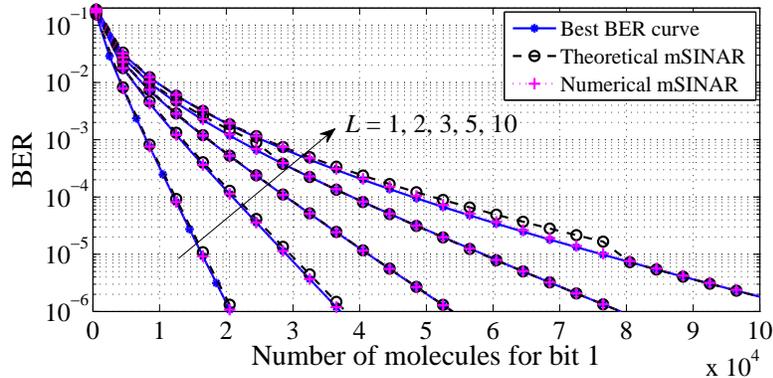}
    \label{Ts_1_verification}
    }
        \subfigure[absorbing receiver with $T_s = 2$]{
	\includegraphics[width=4.5in]{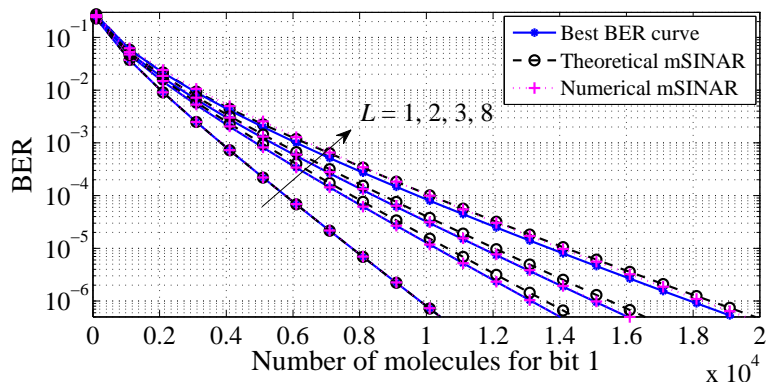}
	\label{Ts_2_verification}
	}
    \caption{BER versus $\cal Q$ performance of the proposed mSINAR: numerical verification, where the passive receiver with ${T_s} = \left\{ {1,2} \right\}$ is considered.}
    \label{Passive_receiver_verification}
\end{figure}

Fig.~\ref{Passive_receiver_optimal_detection_interval_stability} depicts the convergence of ${\left[ {{n_1},{n_2}} \right]^*}$ for the passive receivers with $T_s = 1$ and $L = 2,3,5,10$. Similar to Fig.~\ref{Absorbing_receiver_optimal_detection_interval_stability}, ${\left[ {{n_1},{n_2}} \right]^*}$ is convergent and its convergence rate is inversely proportional to $L$. Moreover, ${\left[ {{n_1},{n_2}} \right]^*}$ obtained from mSINAR almost perfectly matches that obtained from the exhaustive search for $\arg \min {P_e}$, compared with Fig.~\ref{Absorbing_receiver_optimal_detection_interval_stability}. This can be interpreted as follows: the passive receiver can be regarded as an absorbing receiver with limited sampling~times, and due to a large sampling distance, it has some fault tolerance for the optimal detection interval.

\subsection{Validation of Theoretical Results}

In this subsection, we verify the validity of the detection interval optimization obtained from Section III. Specifically, the simulated ${\left[ {{t_1},{t_2}} \right]^*}$ (or ${\left[ {{n_1},{n_2}} \right]^*}$) obtained from \eqref{mSINAR_objective} (or \eqref{detection_interval_passive}) and the theoretical counterpart obtained from \textbf{\emph{Propositions 1-3}} (or \textbf{\emph{Proposition 4}}) are involved, respectively, corresponding to ``Numerical mSINAR" and ``Theoretical mSINAR". For reference, we also include the best BER curve obtained from $\mathop {\arg \min } {P_e}$ via the exhaustive search.

Fig. \ref{Absorbing_receiver_verification} shows the resulting BER versus $\cal Q$ performance, where the absorbing receiver with ${T_s} = \left\{ {0.2,0.3} \right\}$ is considered. It can be seen from Fig.~\ref{Ts_0_2_verification} that the BER curve from the numerical mSINAR perfectly matches the best BER curve for all considered $L$. Then, as $T_s$ increases from $0.2$ to $0.3$, a similar phenomenon is also described in Fig.~\ref{Ts_0_3_verification}. However, compared with the perfect match in Fig.~\ref{Ts_0_2_verification}, we can observe that there is a slight gap between the above BER curves in Fig.~\ref{Ts_0_3_verification}, especially for $L = 8$. This is because the denominator of \eqref{mSINAR} (i.e., the interference) decreases with an increasing $T_s$, thus accelerating the convergence rate of mSINAR. Therefore, we can predict that as $\cal Q$ increases, the BER with the simulated ${\left[ {{t_1},{t_2}} \right]^*}$ approaches the best BER. The above facts indicate that the proposed mSINAR can be good alternatives to measure the BER performance for the absorbing receiver. Moreover, as can be observed from Fig. \ref{Absorbing_receiver_verification}, the BER with the derived ${\left[ {{t_1},{t_2}} \right]^*}$ is asymptotically tight with that of the other two cases for most $\cal Q$, verifying the accuracy of the theoretical ${\left[ {{t_1},{t_2}} \right]^*}$. On the other hand, we also find that for a few $\cal Q$, the BER with the theoretical ${\left[ {{t_1},{t_2}} \right]^*}$ cannot agree well with that of the other two cases, especially
when $L=10$ and $T_s=0.2$. This is attributed to the fact: $\cal Q$ corresponding to the most obvious BER mismatch among all curves is $\mathcal {\hat Q}$ defined in \eqref{mSINAR_objective}; whereas when $0 < {\cal Q} < \hat {\cal Q}$, the derivation of theoretical ${\left[ {{t_1},{t_2}} \right]^*}$ neglects the impact of the noise, thus leading to this mismatch.

Fig.~\ref{Passive_receiver_verification} illustrates the resulting BER versus $\cal Q$ performance, where the passive receivers with $T_s = 1$ and~$2$ are considered in Fig.~\ref{Ts_1_verification} and Fig.~\ref{Ts_2_verification}, respectively. Similar to the absorbing receiver, one can easily observe from Fig. \ref{Passive_receiver_verification} that the BER curves corresponding to the theoretical mSINAR agree with that of the numerical counterparts and the best BER curves approximately, showing the accuracy of the derived $ {\left[ {{n_1},{n_2}} \right]^*}$ and the effectiveness of the mSINAR metric for the passive receiver. Moreover, we also can find that the BER mismatch near the critical point $\hat {\cal Q}$ will be more pronounced than that of the absorbing receiver. This can be accounted for by the fact that the passive receiver itself cannot reduce ISI via absorbing the information molecules, thereby increasing the proportion of the counting noise in the received signals and further weakening the effectiveness using SID to approximately calculate $ {\left[ {{n_1},{n_2}} \right]^*}$ when $0 < {\cal Q} < \hat {\cal Q}$.

\subsection{Performance of The Proposed mSINAR}

\begin{figure}[t]
    \centering
    \subfigure[absorbing receiver with $T_s = 0.2$]{
        \includegraphics[width=4.5in]{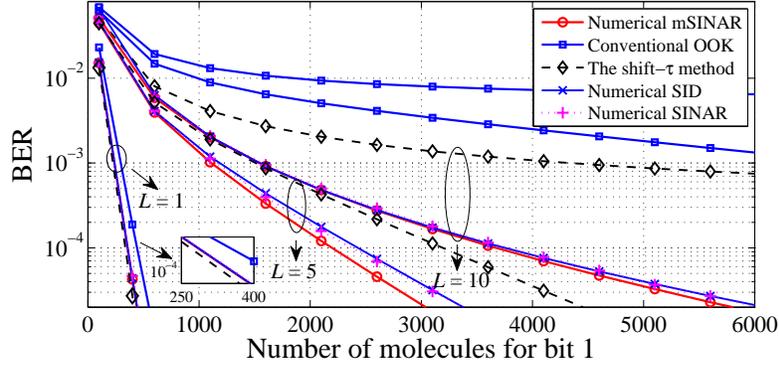}
    \label{Comparison_Ts_0_2}
    }
        \subfigure[absorbing receiver with $T_s = 0.3$]{
	\includegraphics[width=4.5in]{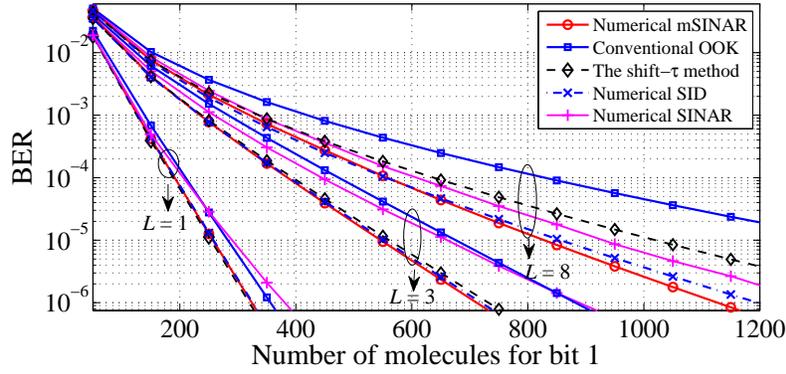}
	\label{Comparison_Ts_0_3}
	}
    \caption{BER comparison among ``numerical mSINAR'', ``numerical SINAR'', ``numerical SID'', ``the shift-$\tau$ method'', and the conventional OOK for the absorbing receiver.}
    \label{Absorbing_receiver_comparison}
\end{figure}

In this subsection, we compare the achievable performance between the proposed mSINAR and other existing schemes. Specifically, based on the concept of optimizing the detection interval during a symbol duration, we include the existing SINAR and SID as benchmarks to estimate the effectiveness of mSINAR, where the detection intervals are obtained from the numerical search for all considered metrics. Besides, to further explore the potential advantage of mSINAR, we also include the shift-$\tau$ method investigated in \cite{The_shift_method_1} and the conventional OOK for comparison. For clarity, in the following figures, we will no longer show the best BER curve, since the BER curves from the numerical mSINAR are almost perfectly matched to the best BER curves in the previous subsection.

\begin{figure}[t]
    \centering
    \subfigure[passive receiver with $T_s = 1$]{
        \includegraphics[width=4.5in]{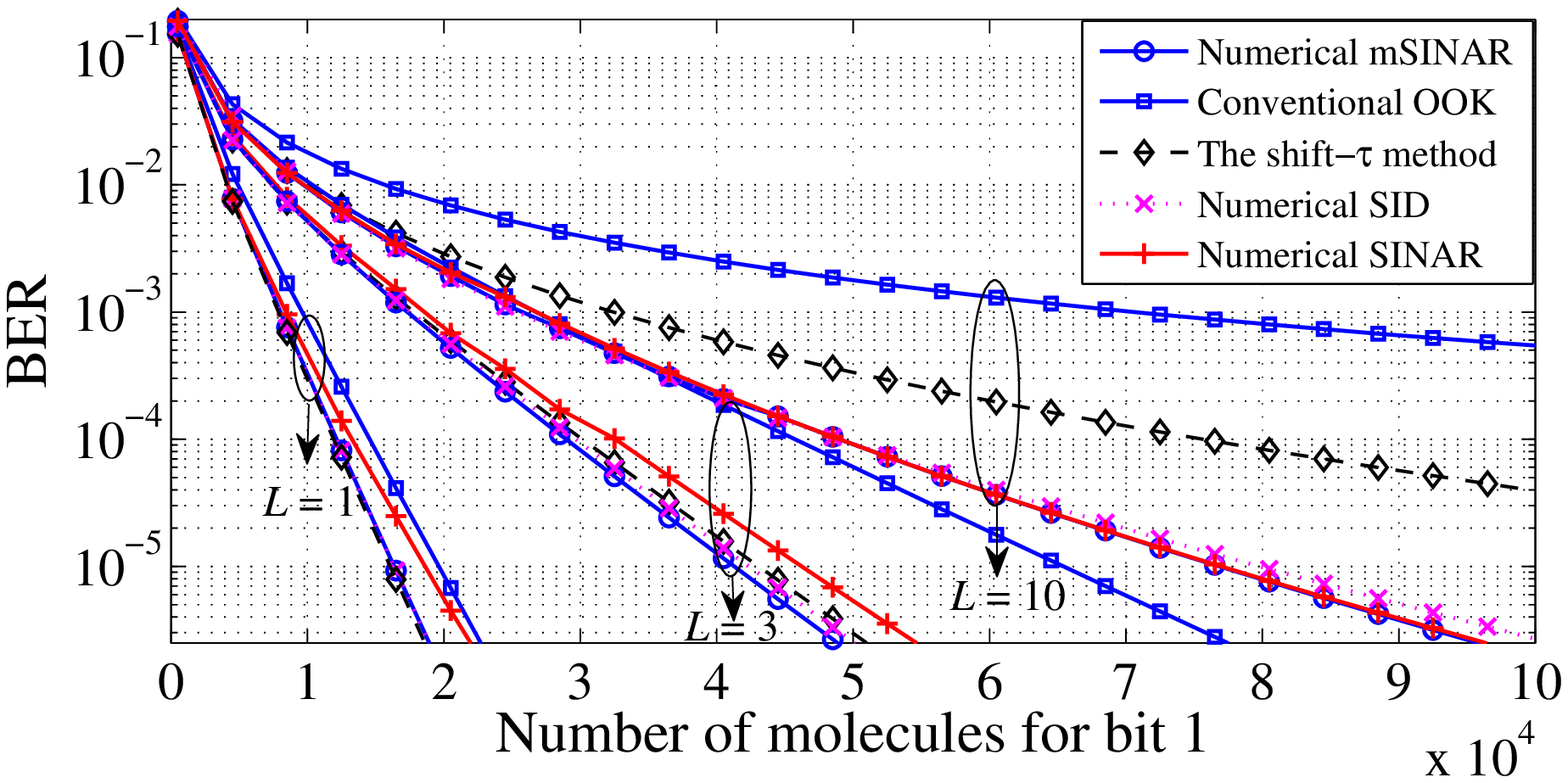}
    \label{comparison_Ts_1}
    }
        \subfigure[passive receiver with $T_s = 2$]{
	\includegraphics[width=4.5in]{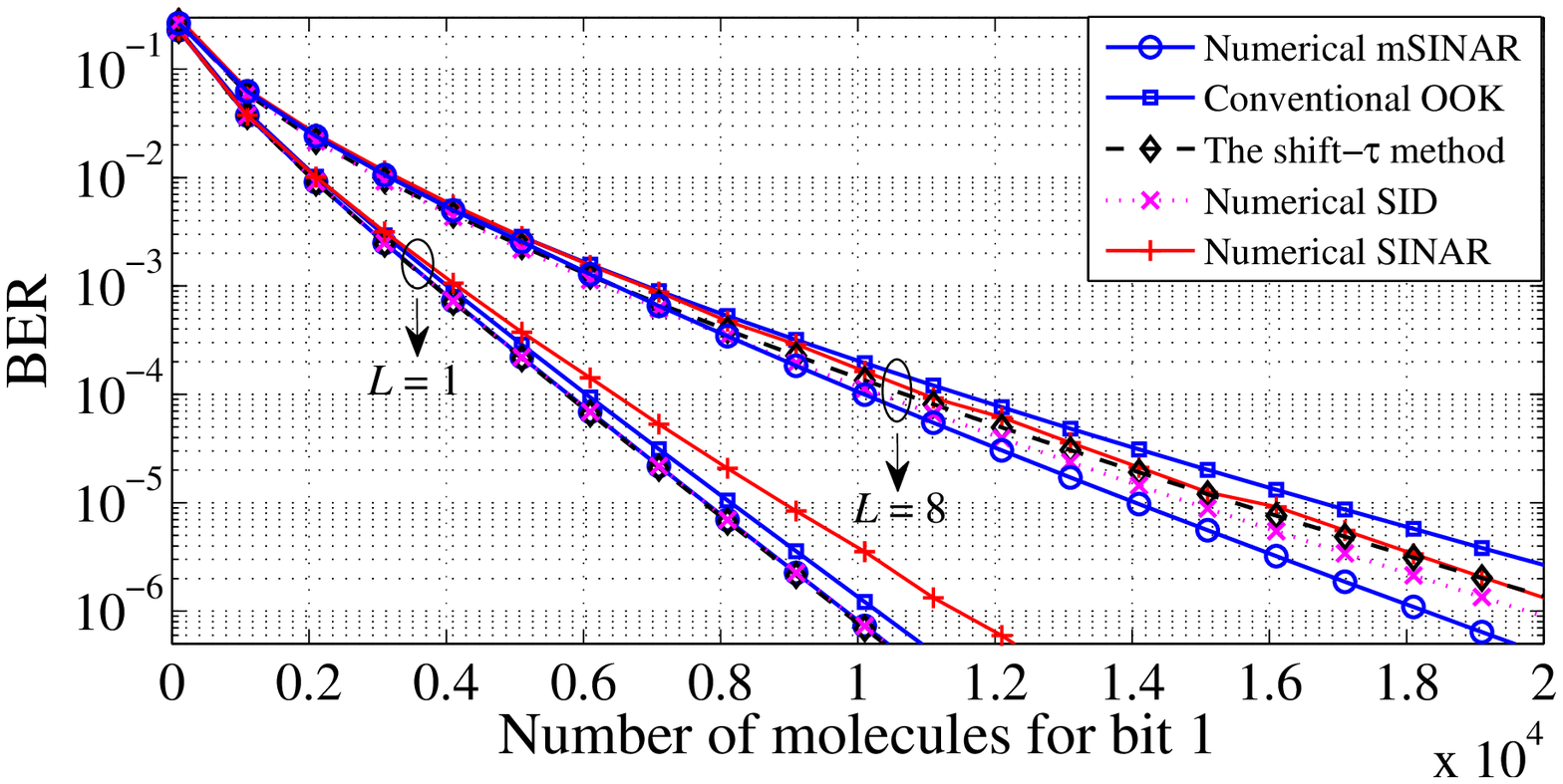}
	\label{Comparison_Ts_2}
	}
    \caption{BER comparison among ``numerical mSINAR'', ``numerical SINAR'', ``numerical SID'', ``the shift-$\tau$ method'', and the conventional OOK for the passive receiver.}
    \label{Passive_receiver_comparison}
\end{figure}

Fig. \ref{Absorbing_receiver_comparison} shows the BER performance comparison among ``numerical mSINAR'', ``numerical SINAR'', ``numerical SID'', ``the shift-$\tau$ method'', and the conventional OOK, where the absorbing receiver with ${T_s} = \left\{ {0.2,0.3} \right\}$ is considered. As expected, both these schemes using the detection interval optimization and the shift-$\tau$ method outperform the conventional OOK cases. Additionally, we can observe from Fig. \ref{Absorbing_receiver_comparison} that the BER curves with mSINAR almost perform best in all considered schemes for $L > 1$; whereas when $L = 1$, the shift-$\tau$ method is slightly superior to the numerical mSINAR, especially in Fig.~\ref{Comparison_Ts_0_2}. This can be interpreted as follows: when $L$ is small and $t > t_{\max}$, the amplitude of the ISI signal cannot exceed that of the expected signal, such that the received information belonging to $\left[ {{T_s},\tau  + {T_s}} \right]$ contributes to signal detection and the shift-$\tau$ method performs better; however, with the increase of $L$, the cumulative interference intensity in $\left[ {{T_s},\tau  + {T_s}} \right]$ gradually rises and finally surpasses the expected signal intensity. Here, since the detection interval is always in the range of the current symbol period, the proposed mSINAR does not bring the extra interference generated in $\left[ {{T_s},\tau  + {T_s}} \right]$ and thus obtain the performance gain. Particularly, besides the existing ISI signal, the transmitted signal of the next symbol duration is also regarded as the interference in $\left[ {{T_s},\tau  + {T_s}} \right]$. Finally, we compare the BER curves using different performance metrics. As for SID, it can be seen from Fig. \ref{Absorbing_receiver_comparison} that the degree of agreement between the BER curve corresponding to SID and that corresponding to mSINAR is not stable, which is caused by SID ignoring the influence of noise. Concerning SIANR, we find that the BER curve using SINAR will gradually deviate from that using mSINAR as $\cal Q$ grows. This arises from the fact that as the variance of the noise in \eqref{SINAR} decreases with an increasing $\cal Q$, all ${\left[ {{t_1},{t_2}} \right]^*}$ derived from SINAR with different $L$ and $T_s$ finally approaches nearby the peak time of the molecule concentration $t_{\max}$, which violates the convergence of the true~${\left[ {{t_1},{t_2}} \right]^*}$.

As for the passive receiver, Fig.~\ref{Passive_receiver_comparison} shows the simulation results similar to Fig.~\ref{Absorbing_receiver_comparison}. By carefully observing Fig. \ref{Absorbing_receiver_comparison}-\ref{Passive_receiver_comparison}, we can find that the achieved gain of the proposed mSINAR over the shift-$\tau$ decreases with an increasing $T_s$ whether the absorbing or passive receiver is considered, since the increase of $T_s$ weakens the amplitude of the ISI signals and thus lowers the extra interference generated in $\left[ {{T_s},\tau  + {T_s}} \right]$. Comparing mSINAR with SID and SINAR, we can see that the latter two cannot provide a stable quantification for BER trends and also cannot derive a close-to-optimal detection interval for all considered cases, proving that mSINAR is more suitable to measure the possible ${\left[ {{t_1},{t_2}} \right]^*}$.

\section{Conclusion} \label{Conclusion}

In this paper, we proposed optimizing the detection interval to mitigate the impact of ISI signals for a typical MCvD system. Constrained by the complex expression of BER, a new performance indicator, namely mSINAR, was proposed to simplify the optimization process. Based on the mSINAR method, the objective function related to the optimal detection interval has been formulated. By decomposing the objective functions as two sub-problems, we derived the closed-form expression of this interval when the absorbing and passive receivers are considered. Moreover, numerical simulations revealed the convergence of the optimal detection interval. Finally, we proved that the proposed mSINAR method significantly outperforms the benchmark schemes when the ISI length is more than 1 and gradually achieves the best BER performance for the considered MCvD system.

\begin{appendices}

\section{}\label{Appendix A}

Assuming $L=1$, \eqref{find_interval_AB} can be expressed as
\begin{align}\label{L_1_AB}
\frac{3}{2}\left( {{T_s} + t_i^*} \right)t_i^*\left\{ {\ln \left( {1 + \frac{{t_i^*}}{{{T_s}}}} \right) - \ln \left( {\frac{{t_i^*}}{{{T_s}}}} \right)} \right\} = {m^2}{T_s},~i=1,2
\end{align}
where $m = \frac{d}{{\sqrt {4D} }}$. For clarity, we first calculate $t_1^*$. It is clear from \eqref{L_1_AB} that $\ln \left(  \cdot  \right)$ is a main obstacle to calculate $t_1^*$. Considering the fact that the information at $t_{\max}$ should be collected for signal detection, we have $t_1^* \in \left( {0,{t_{\max }}} \right)$ and $t_2^* \in \left( {{t_{\max }},{T_s}} \right]$. Besides, compared with $T_s$, $t_{\max}$, the time of peak concentration, is generally small. Thereby, we can find that $\frac{{{t_1^*}}}{{{T_s}}}$ is much closer to 0 than 1. Here, we can use the approximation methods to simplify \eqref{L_1_AB}. Specifically, the $\left[ {1,1} \right]$ Padé approximation of $\ln (1+x)$ can be written as $\ln (1 + x) \approx \frac{{2x}}{{2 + x}}$, and $\ln (x)$ can be approximately expressed as $\ln x = \frac{{20\left( {x - 1} \right)}}{{7 + 15x}}$ \cite{The_shift_method_1}. Substituting the above approximations into \eqref{L_1_AB} yields
\begin{align}\label{L_1_AB_Pade}
\hspace{-0.2cm}29{\left( {t_1^*} \right)^4} + 23{\left( {t_1^*} \right)^3}{T_s} + {\left( {t_1^*} \right)^2}{T_s}\left( {34{T_s} - 10{m^2}{T_s}} \right)
+ t_1^*T_s^2\left( {40{T_s} - \frac{{74}}{3}{m^2}} \right) - \frac{{28}}{3}{m^2}T_s^3 = 0.
\end{align}
Considering that ${t_1^*}$ should be a small value and $0 < {t_1^*} < {t_{\max }}$, ${t_1^*}$ terms whose order is higher than one can be neglected. Thus, \eqref{L_1_AB_Pade} can be approximately solved as
\begin{align}\label{L_1_AB_final}
t_1^* \approx \frac{{28{m^2}{T_s}}}{{120{T_s} - 74{m^2}}}.
\end{align}
Moreover, given $t_2^* \in \left( {{t_{\max }},{T_s}} \right]$, we can find that $h\left( {t_2^*} \right) - h\left( {{T_s} + t_2^*} \right) > 0$ almost always holds in \eqref{find_interval_AB} if $L=1$. Therefore, we have $t_2^* = T_s$. In summary, we can obtain
\begin{align}\label{Final_optimal_interval_AB_appendix}
{\left[ {{t_1},{t_2}} \right]^*} \approx \left[ {\frac{{28{m^2}{T_s}}}{{120{T_s} - 74{m^2}}},{T_s}} \right].
\end{align}

\vspace{0.3cm}
\section{}\label{Appendix B}

Assuming $L>1$, \eqref{find_interval_AB} can be re-written as
\begin{align}\label{L>1_AB}
\frac{1}{{\sqrt {t{{_i^*}^3}} }}\exp \left( { - \frac{{{m^2}}}{{t_i^*}}} \right) = \sum\limits_{k = 1}^L {\frac{1}{{\sqrt {{{\left( {k{T_s} + t_i^*} \right)}^3}} }}\exp \left( { - \frac{{{m^2}}}{{\left( {k{T_s} + t_i^*} \right)}}} \right)} , i = 1,2 .
\end{align}
For clarity, we first calculate $t_1^*$. Considering $t_1^* \in \left( {0,{t_{\max }}} \right)$ and $t_{\max} << T_s$, we approximately have
\begin{align}
\exp \left( { - \frac{{{m^2}}}{{\left( {k{T_s} + t_1^*} \right)}}} \right) \approx \exp \left( { - \frac{{{m^2}}}{{\left( {k{T_s} + \hat t_1^*} \right)}}} \right) \nonumber,~k= 1, \cdots, L
\end{align}
where $\hat t_1^*$ denotes the optimal $t_1$ when $L = 1$. Moreover, with the aid of the first order of Taylor expression of ${\left( {1 + x} \right)^a}$, \eqref{L>1_AB} can be expressed~as
\begin{align}\label{L>1_AB_simple_t1}
% %  {g^{-\frac{3}{2}}}\exp \left( { - \frac{{{m^2}}}{{{T_sg}}}} \right) \approx \sum\limits_{k = 1}^L {{a_k}\left( {1 - \frac{3}{2k}g} \right)},
% {g^{\frac{5}{2}}}\left\{ {\sum\limits_{k = 1}^L {{a_k}\left( {1 - \frac{3}{{2k}}} \right)} } \right\} - {g^{\frac{3}{2}}}\left\{ {\sum\limits_{k = 1}^L {{a_k}} } \right\} + \exp \left( { - \frac{{{m^2}}}{{{T_s}}}{g^{ - 1}}} \right) \approx 0,
{\left( {\frac{{t_1^*}}{{{T_s}}}} \right)^{\frac{5}{2}}}\left\{ {\sum\limits_{k = 1}^L {{a_k}\left( {1 - \frac{3}{{2k}}} \right)} } \right\} - {\left( {\frac{{t_1^*}}{{{T_s}}}} \right)^{\frac{3}{2}}}\left\{ {\sum\limits_{k = 1}^L {{a_k}} } \right\} + \exp \left( { - \frac{{{m^2}}}{{t_1^*}}} \right) \approx 0,
\end{align}
where ${a_k} = \frac{1}{{\sqrt {{k^3}} }}\exp \left( { - \frac{{{m^2}}}{{\left( {k{T_s} + \hat t_1^*} \right)}}} \right)$. However, it can be seen from \eqref{L>1_AB_simple_t1} that we hardly obtain a closed-form solution of  $t_1^*$. The reason can be described as follows: 1) due to $\frac{{{m^2}}}{{t_1^*}} \ge 1.5$, $\exp {\left( { - \frac{{{m^2}}}{{t_1^*}}} \right)}$ cannot be simplified well by Taylor series or other approximation methods; 2) there exist the fractional power exponents ${\left( {\frac{{t_1^*}}{{{T_s}}}} \right)^{\frac{5}{2}}}$ and ${\left( {\frac{{t_1^*}}{{{T_s}}}} \right)^{\frac{3}{2}}}$ in \eqref{L>1_AB_simple_t1}, which are hard to deal with. Therefore, in order to obtain a better expression for $t_1^*$, we need to simplify the summation caused by the $L$ ISI signals in \eqref{L>1_AB}. First, given that the ISI from the last time slot plays a dominant role in all ISI signals, it is reasonable to assume $t_1^* \approx \hat t_1^*$. Second, one can easily observe from the property of the diffusion channel that $h\left(t\right)$ decreases slowly when $t$ is relatively large. As per the above, we can approximate the ratio of ${h\left( {k{T_s} + t} \right)}$ with $k \ge 2$ to ${h\left(T_s + t \right)}$ at $t_1^*$ as the ratio at $\hat t_1^*$, mathematically expressed as
\begin{align}\label{ratio_AB}
\frac{{h\left( {k{T_s} + t_1^*} \right)}}{{h\left( {{T_s} + t_1^*} \right)}} \approx \frac{{h\left( {k{T_s} + \hat t_1^*} \right)}}{{h\left( {{T_s} + \hat t_1^*} \right)}}.
\end{align}
Substituting \eqref{ratio_AB} into \eqref{L>1_AB} yields
\begin{align}\label{L>1_AB_approximate}
\frac{1}{{\sqrt {t{{_i^*}^3}} }}\exp \left( { - \frac{{{m^2}}}{{t_1^*}}} \right) = \frac{1}{{\sqrt {{{\left( {{T_s} + t_1^*} \right)}^3}} }}\exp \left( { - \frac{{{m^2}}}{{\left( {{T_s} + t_1^*} \right)}}} \right)\sum\limits_{k = 1}^L {\left\{ {\frac{{h\left( {k{T_s} + \hat t_1^*} \right)}}{{h\left( {{T_s} + \hat t_1^*} \right)}}} \right\}} .
\end{align}
Let us define $\mathcal{I} = \sum\limits_{k = 1}^L {\left\{ {\frac{{h\left( {k{T_s} + \hat t_1^*} \right)}}{{h\left( {{T_s} + \hat t_1^*} \right)}}} \right\}} $, \eqref{L>1_AB_approximate} can be further simplified as
\begin{align}\label{L>1_AB_approximate_simple_1}
{m^2}{T_s} = \frac{3}{2}t_1^*\left( {{T_s} + t_1^*} \right)\left( {\ln \left( {1 + \frac{{t_1^*}}{{{T_s}}}} \right) - \ln \left( {\frac{{t_1^*}}{{{T_s}}}} \right)} \right) - t_1^*\left( {{T_s} + t_1^*} \right)\ln {\cal I}.
\end{align}
With the aid of the approximations of $\ln \left( {1 + x} \right)$ and $\ln \left( {x} \right)$, we can approximate \eqref{L>1_AB_approximate_simple_1} as
\begin{align}\label{L>1_AB_approximate_simple_2}
&{\left( {t_1^*} \right)^4}\left( {29 - 10\ln {\cal I}} \right) + {\left( {t_1^*} \right)^3}{T_s}\left( {23 - \frac{{104}}{3}\ln {\cal I}} \right)+ {\left( {t_1^*} \right)^2}{T_s}\left( {34{T_s} - 34{T_s}\ln {\cal I} - 10{m^2}{T_s}} \right) \nonumber \\
&+ t_1^*T_s^2\left( {40{T_s} - \frac{{28}}{3}{T_s}\ln {\cal I} - \frac{{74}}{3}{m^2}} \right) - \frac{{28}}{3}{m^2}T_s^3 = 0.
\end{align}
When ${t_1^*}$ terms whose order is higher than one are neglected, ${t_1^*}$ can be solved as
\begin{align}\label{L>1_AB_approximate_solution_t1}
t_1^* \approx \frac{{28{m^2}{T_s}}}{{120{T_s} - 28{T_s}\ln {\cal I} - 74{m^2}}}.
\end{align}
Next, we focus on the calculation of $t_2^*$. Provided that when $L$ is not enough large, the ISI signals cannot exceed the expected signal. Therefore, we have if
\begin{align}\label{condition_t2_not_TS}
1 - \sum\limits_{k = 1}^L {\frac{1}{{\sqrt {{{\left( {1 + k} \right)}^3}} }}} \exp \left( {\frac{k}{{\left( {1 + k} \right)}}\frac{{{m^2}}}{{{T_s}}}} \right) \ge 0
\end{align}
holds, $t_2^* = {T_s}$. On the other hand, when \eqref{condition_t2_not_TS} cannot be satisfied, the calculation procedure of $t_2^*$ is similar to that of $t_1^*$. The first step is still to simplify \eqref{L>1_AB}. Similar to \eqref{ratio_AB}, we also have the following expression to approximate
the ratio of ${h\left( {k{T_s} + t} \right)}$ with $k \ge 2$ to ${h\left(T_s + t \right)}$ at $t_2^*$ as the ratio at $\hat t_2^*$,
\begin{align}\label{ratio_AB_t2}
\frac{{h\left( {k{T_s} + t_2^*} \right)}}{{h\left( {{T_s} + t_2^*} \right)}} \approx \frac{{h\left( {k{T_s} + \hat t_2^*} \right)}}{{h\left( {{T_s} + \hat t_2^*} \right)}}.
\end{align}
Particularly, we set the pre-defined $\hat t_2^*$ as the mean for possible $t_2$, i.e., $\hat t_2^* = \frac{1}{2}\left( {{t_{\max }} + {T_s}} \right)$. This is because, compared with the assumption of $t_1^* \approx \hat t_1^*$, we cannot identify the impact of the last ISI signal from $t_2^* = {T_s}$ when $L=1$; and the fluctuation of all signals may be much sharply due to the larger $\left( {{t_{\max }},{T_s}} \right]$. Per the aforementioned, \eqref{L>1_AB} can be re-written as
% \begin{align}\label{L>1_AB_approximate_t2}
% {\left( {1 + \frac{{{T_s}}}{{t_2^*}}} \right)^{ - \frac{3}{2}}}\exp \left( { - \frac{{{m^2}}}{{\left( {{T_s} + t_2^*} \right)}}} \right)\sum\limits_{k = 1}^L {\frac{{h\left( {k{T_s} + \hat t_2^*} \right)}}{{h\left( {{T_s} + \hat t_2^*} \right)}}}  - \exp \left( { - \frac{{{m^2}}}{{t_2^*}}} \right) \approx 0.
% \end{align}
% Defining ${\cal V} = \sum\limits_{k = 1}^L {\frac{{h\left( {k{T_s} + \hat t_2^*} \right)}}{{h\left( {{T_s} + \hat t_2^*} \right)}}} $, \eqref{L>1_AB_approximate_t2} can be simplified as
\begin{align}\label{L>1_AB_more_approximate_t2_1}
\frac{3}{2}t_2^*\left( {{T_s} + t_2^*} \right)\left( {\ln \left( {1 + \frac{{t_2^*}}{{{T_s}}}} \right) - \ln \left( {\frac{{t_2^*}}{{{T_s}}}} \right)} \right) - t_2^*\left( {{T_s} + t_2^*} \right)\ln {\cal V} ={m^2}{T_s} ,
\end{align}
where ${\cal V} = \sum\limits_{k = 1}^L {\frac{{h\left( {k{T_s} + \hat t_2^*} \right)}}{{h\left( {{T_s} + \hat t_2^*} \right)}}} $.
It is worth noting that that ${\frac{{t_2^*}}{{{T_s}}}}$ is much closer 1 than~0. Hence, we have $\ln (1 + x) \approx \frac{{2x}}{{2 + x}}$ and $\ln (x) \approx \frac{{2\left( {x - 1} \right)}}{{1 + x}}$ for $x \to 1$. Here, \eqref{L>1_AB_more_approximate_t2_1} can be rewritten as
\begin{align}\label{L>1_AB_more_approximate_t2_2}
{\left( {\frac{{t_2^*}}{{{T_s}}}} \right)^3}\ln {\cal V} + 3{\left( {\frac{{t_2^*}}{{{T_s}}}} \right)^2}\ln {\cal V} + \frac{{t_2^*}}{{{T_s}}}\left( {2\ln {\cal V} + \frac{{{m^2}}}{{{T_s}}} - 6} \right) + \frac{{{2m^2}}}{{{T_s}}} \approx 0.
\end{align}
In \eqref{L>1_AB_more_approximate_t2_2}, one of the three roots, which needs to be solved, is in the range of $ {t_{\max }} < {t_2^*} < {T_s}$. To obtain an effective solution, we need to explore the impact of the discriminant of this cubic equation. Define the discriminant of \eqref{L>1_AB_more_approximate_t2_2} as
\begin{align}\label{discriminant_of_cubic}
{\Delta _1} = {\left( {\ln {\mathcal {V}}} \right)^2}\left( {{\left( {3\gamma  - \frac{{18{m^2}}}{{{T_s}}}} \right)^2} - \left( {\frac{{12\gamma }}{{\ln {\mathcal {V}}}} - 36} \right)\left( {{\gamma ^2} - \frac{{18{m^2}\ln {\mathcal {V}}}}{{{T_s}}}} \right)} \right),
\end{align}
where $\gamma  = \left( {2\ln {\mathcal {V}} + \frac{{{m^2}}}{{{T_s}}} - 6} \right)$. Then, from the first derivative of ${\Delta _1}$, we can find that when ${\mathcal {V}} < \exp\left(\frac{{5{m^2}}}{{4{T_s}}} + \frac{3}{2}\right)$
% \begin{align}\label{Delta_1_decreasing}
%  {\mathcal {V}} < \exp\left(\frac{{5{m^2}}}{{4{T_s}}} + \frac{3}{2}\right) \nonumber
% \end{align}
can be satisfied, ${\Delta _1}$ is an increasing function with respect to $\ln {\mathcal {V}}$ or $L$. As can be seen from the definition of $ {\mathcal V}$, the required condition is easy to achieve. Then, we can find that when $\ln {\cal V} \to 0$ (i.e., $L \to 1$), the initial value of ${\Delta_1}$ is negative, mathematically expressed~as
\begin{align}
{\Delta _1} =  - \left( {72 - \frac{{12{m^2}}}{{{T_s}}}} \right){\left( {\frac{{{m^2}}}{{{T_s}}} - 6} \right)^2}\ln {\cal V} + {\cal O}\left( {{{\left( {\ln {\cal V}} \right)}^2}} \right) + {\cal O}\left( {{{\left( {\ln {\cal V}} \right)}^3}} \right) + {\cal O}\left( {{{\left( {\ln {\cal V}} \right)}^4}} \right) < 0. \nonumber
\end{align}
Therefore, if $\Delta_1<0$ can be satisfied, \eqref{L>1_AB_more_approximate_t2_2} can be solved as
\begin{align}\label{t2_approximate_solution_1}
t_2^* \approx   \frac{{-{3 + \left( {\sqrt[3]{{{s_1}}} + \sqrt[3]{{{s_2}}}} \right)} }}{3}{T_s},
\end{align}
where
\begin{align}
{s_i} =   - \left( {\frac{{81}}{{\ln {\mathcal {V}} }} + \frac{{27{m^2}}}{{2{T_s}\ln {\mathcal {V}} }}} \right) + {\left( { - 1} \right)^{i - 1}}\frac{{\sqrt {9{\Delta _1}} }}{{2{{\left( {\ln {\mathcal {V}} } \right)}^2}}},~i = 1,2.\nonumber
\end{align}
As $L$ goes large until $\Delta_1 \ge 0$, more ISI signals are introduced and $t_2^*$ gradually decreases. At this moment, we decide to neglect ${\frac{{t_2^*}}{{{T_s}}}}$ with the highest order and then \eqref{discriminant_of_cubic} can be transformed into the quadratic equation. Let us define the discriminant of the updated quadratic equation as
\begin{align}\label{Delta_2}
{\Delta_2} = {\left( {\frac{{{m^2}}}{{{T_s}}} - 6} \right)^2} + 4{\left( {\ln {\cal V}} \right)^2} - \left( {\frac{{20{m^2}}}{{{T_s}}} + 24} \right)\ln {\cal V}.
\end{align}
Taking the first derivative of ${\Delta _2}$ with respect to ${\cal V}$ yields
\begin{align}
\frac{{\partial {\Delta _2}}}{{\partial \ln {\cal V}}} = 8\ln {\cal V} - \frac{{20{m^2}}}{{{T_s}}} - 24 < 0, \nonumber
\end{align}
suggesting that ${\Delta _2}$ is degraded with the growth of $L$. Moreover, it is clear from \eqref{Delta_2} that when $\ln {\cal V}$ is not large enough, it is possible for $\Delta_2 \ge 0$. Thus, when $\Delta_1 \ge 0$ and $\Delta_2 \ge 0$, \eqref{L>1_AB_more_approximate_t2_2} can be computed as
\begin{align}\label{t2_approximate_solution_2}
t_2^* \approx \frac{{ - \gamma  + \sqrt {{\Delta _2}} }}{6{\ln {\mathcal {V}} }}{T_s},
\end{align}
Furthermore, when $\Delta_1 \ge 0$ and $\Delta_2 < 0$, we approximate $t_2^*$ as
\begin{align}\label{t2_approximate_solution_3}
t_2^* \approx -\frac{{ \gamma }}{6{\ln {\mathcal {V}} }}{T_s}.
\end{align}
So far, the calculation for ${\left[ {{t_1},{t_2}} \right]^*}$ has been finished and the approximate solutions of ${\left[ {{t_1},{t_2}} \right]^*}$ for all possible $L$ can be found in \textbf{\emph{Proposition~2}}.

\vspace{0.3cm}
\section{}\label{Appendix C}

According to \eqref{find_interval_passive_simple}, we first investigate the impact of the strongest ISI tap, i.e., assuming $L=1$. Since the expected signal always overwhelms the ISI signal when $t>t_{\max}$, we have $n_2^* = N$. Here, \eqref{find_interval_passive_simple} can be re-written as
\begin{align}\label{find_interval_passive_L_1}
n_1^*{t_s}\left( {n_1^*{t_s} + {T_s}} \right) + {{\hat m}^2}{T_s} = \frac{3}{2}n_1^*{t_s}\left( {n_1^*{t_s} + {T_s}} \right)\ln \frac{{\left( {n_1^*{t_s} + {T_s}} \right)}}{{n_1^*{t_s}}},
\end{align}
where $\hat m = \frac{{d + r}}{{\sqrt {4D} }}$. At this point, with the aid of the approximation again, we can conclude
\begin{align}\label{Final_optimal_interval_Passive_L_1_appendix}
{\left[ {{n_1},{n_2}} \right]^*} \approx \left[ {\left\lceil {\frac{{28{{\hat m}^2}{T_s}}}{{\left( {120{T_s} -  74{{\hat m}^2}} \right){t_s}}}} \right\rceil,N} \right].
\end{align}
Next, we try to find a feasible solution for ${\left[ {{n_1},{n_2}} \right]^*}$ when $L>1$. It is clear from \eqref{find_interval_passive_simple} that the summation of $L$ ISI signals and $\exp \left(  \cdot  \right)$ is a huge challenge to calculate ${\left[ {{n_1},{n_2}} \right]^*}$. Therefore, in the following, we also use a similar method to the absorbing receiver to apply the last ISI to approximate the remaining ISI signal.

For clarity, we first calculate $n_1^*$. It is assumed that the ratio between ${p_{n,1}}$ and ${p_{n,k}}$ at $n_1^*$ can be approximated as the ratio at $\hat n_1^*$ for $k = 2, \cdots ,L$, i.e.,
\begin{align}\label{ratio_passive_n1}
\frac{{{p_{n_1^*,k}}}}{{{p_{n_1^*,1}}}} \approx \frac{{{p_{\hat n_1^*,k}}}}{{{p_{\hat n_1^*,1}}}},
\end{align}
where $\hat n_1^*$ is assumed to be the optimal first sample when $L=1$, i.e., $\hat n_1^* = \left\lceil {\frac{{28{{\hat m}^2}{T_s}}}{{\left( {120{T_s} - 74{{\hat m}^2}} \right){t_s}}}} \right\rceil$. Substituting \eqref{ratio_passive_n1} into \eqref{find_interval_passive_simple} yields
\begin{align}\label{find_interval_passive_simple_n1}
\frac{1}{{{{\left( {n_1^*{t_s} + {T_s}} \right)}^{3/2}}}}\exp \left( { - \frac{{{{\hat m}^2}}}{{\left( {n_1^*{t_s} + {T_s}} \right)}}} \right)\sum\limits_{k = 1}^L {\frac{{{p_{\hat n_1^*,k}}}}{{{p_{\hat n_1^*,1}}}}}  = \frac{1}{{{{\left( {n_1^*{t_s}} \right)}^{3/2}}}}\exp \left( { - \frac{{{{\hat m}^2}}}{{n_1^*{t_s}}}} \right).
\end{align}
Define ${\cal W} = \sum\limits_{k = 1}^L {{\frac{{{p_{\hat n_1^*,k}}}}{{{p_{\hat n_1^*,1}}}}} } $, \eqref{find_interval_passive_simple_n1} can be further simplified as
\begin{align}
\hspace{-0.2cm}\frac{3}{2}\left( {n_1^*{t_s} + {T_s}} \right)\left( {n_1^*{t_s}} \right)\left( {\ln \left( {1 + \frac{{n_1^*{t_s}}}{{{T_s}}}} \right) - \ln \left( {\frac{{n_1^*{t_s}}}{{{T_s}}}} \right)} \right) - \left( {n_1^*{t_s} + {T_s}} \right)\left( {n_1^*{t_s}} \right)\ln {\cal W} = {{\hat m}^2}{T_s}.
\end{align}
With the approximation of $\ln \left( {1 + x} \right)$ and $\ln \left( { x} \right)$, $n_1^*$ can be solved as
\begin{align}
n_1^* \approx \left\lceil {\frac{{28{{\hat m}^2}{T_s}}}{{\left( {120{T_s} - 28{T_s}\ln {\cal W} - 74{{\hat m}^2}} \right){t_s}}}} \right\rceil .
\end{align}

As for $n_2$, we also first obtain the approximation of the ratio between ${p_{n,1}}$ and ${p_{n,k}}$ at $n_2^*$, i.e.,
\begin{align}\label{ratio_passive_n2}
\frac{{{p_{n_2^*,k}}}}{{{p_{n_2^*,1}}}} \approx \frac{{{p_{\hat n_2^*,k}}}}{{{p_{\hat n_2^*,1}}}},
\end{align}
where $k = 2, \cdots ,L$ and $\hat n_2^* = \left\lceil {\frac{{{t_{\max }} + {T_s}}}{{2{t_s}}}} \right\rceil $. By substituting \eqref{ratio_passive_n2} into \eqref{find_interval_passive_simple}, we can obtain
\begin{align}\label{find_interval_passive_simple_n2}
\frac{3}{2}\left( {n_2^*{t_s} + {T_s}} \right)\left( {n_2^*{t_s}} \right)\left( {\ln \left( {1 + \frac{{n_2^*{t_s}}}{{{T_s}}}} \right) - \ln \left( {\frac{{n_2^*{t_s}}}{{{T_s}}}} \right)} \right) - \left( {n_2^*{t_s} + {T_s}} \right)\left( {n_2^*{t_s}} \right)\ln {\cal A} = {{\hat m}^2}{T_s},
\end{align}
where ${\cal A}= \sum\limits_{k = 1}^L {\frac{{{p_{\hat n_2^*,k}}}}{{{p_{\hat n_2^*,1}}}}}$. Similar to \eqref{L>1_AB_more_approximate_t2_2}-\eqref{t2_approximate_solution_3}, we also can solve $n_2^*$. For clarity, ${\left[ {{n_1},{n_2}} \right]^*} $ can be summarized in \eqref{optimal_n1_n2_SID}, where
\vspace{-0.3cm}
\begin{align}
    \hat \gamma  = 2\ln {\cal A} + \frac{{{{\hat m}^2}}}{{{T_s}}} - 6, \nonumber
\end{align}
\vspace{-0.66cm}
\begin{align}
{{\hat s}_i} =  - \left( {\frac{{81}}{{\ln {\cal A}}} + \frac{{27{{\hat m}^2}}}{{2{T_s}\ln {\cal A}}}} \right) + {\left( { - 1} \right)^{i - 1}}\frac{{\sqrt {9{{\hat \Delta }_1}} }}{{2{{\left( {\ln {\cal A}} \right)}^2}}},i = 1,2.\nonumber
\end{align}
Besides, ${{\hat \Delta }_1}$ and ${{\hat \Delta }_2}$ denote the discriminant of the cubic and quadratic equations generated from \eqref{find_interval_passive_simple_n2}, respectively, expressed as
\begin{align}
{{\hat \Delta }_1} = {\left( {\ln {\cal A}} \right)^2}\left( {{{\left( {3\hat \gamma  - \frac{{18{{\hat m}^2}}}{{{T_s}}}} \right)}^2} - \left( {\frac{{12\hat \gamma }}{{\ln {\cal A}}} - 36} \right)\left( {{{\hat \gamma }^2} - \frac{{18{{\hat m}^2}\ln \hat \gamma }}{{{T_s}}}} \right)} \right), \nonumber
\end{align}
\vspace{-0.3cm}
\begin{align}
{{\hat \Delta }_2} = {\left( {\frac{{{{\hat m}^2}}}{{{T_s}}} - 6} \right)^2} + 4{\left( {\ln {\cal A}} \right)^2} - \left( {\frac{{20{{\hat m}^2}}}{{{T_s}}} + 24} \right)\ln {\cal A} \nonumber.
\end{align}
At this point, the theoretical ${\left[ {{n_1},{n_2}} \right]^*}$ has been derived, as shown in \eqref{Final_optimal_interval_Passive_L_1} and \eqref{optimal_n1_n2_SID}.

\end{appendices}

\bibliographystyle{IEEEtran}
\bibliography{IEEEabrv,J_VERSION_Appendix}

\end{document}